%% file: main.tex
\documentclass[%
reprint,
 amsmath,amssymb,
 aps, 
 physrev,
]{revtex4-2}
\usepackage[utf8]{inputenc}
\usepackage[T1]{fontenc}
\usepackage{csquotes}
\usepackage{graphicx}% Include figure files
\usepackage{dcolumn}% Align table columns on decimal point
\usepackage{bm}% bold math
\usepackage{dsfont}
\usepackage{changes}
\usepackage{hyperref}% add hypertext capabilities
\input{pretex}

\input{opdef}

\newcommand{\locc}{\mathrm{LOCC}}

\graphicspath{{figures/}{../figures/}}
\begin{document}

% \preprint{APS/123-QED}

\title{Optimizing LOCC Protocols on Product Stiefel Manifold}% 

\author{Ze-Tong Li}
\affiliation{Thrust of Artificial Intelligence, Information Hub, The Hong Kong University of Science and Technology (Guangzhou), Guangdong 511453, China}
\author{Chengkai Zhu}
\affiliation{Thrust of Artificial Intelligence, Information Hub, The Hong Kong University of Science and Technology (Guangzhou), Guangdong 511453, China}
\author{Xin Wang}%
\email{felixxinwang@hkust-gz.edu.cn}
\affiliation{Thrust of Artificial Intelligence, Information Hub, The Hong Kong University of Science and Technology (Guangzhou), Guangdong 511453, China}

\date{\today}% It is always \today, today,

\begin{abstract}
Characterizing the operational limits of Local Operations and Classical Communication (LOCC) is a central problem in distributed quantum information, yet remains computationally intractable due to the non-convex geometry of the LOCC set. We introduce a geometric framework that embeds the physical constraints of fixed-round LOCC protocols onto the product Stiefel manifold, converting a constrained protocol-design problem into unconstrained Riemannian optimization. We demonstrate this framework through entanglement distillation: by directly optimizing finite-copy LOCC protocols, we discover achievable protocols whose fidelities match positive partial transpose (PPT) upper bounds to within numerical precision, and we provide numerical evidence for both the operational advantage of adaptive communication rounds and the super-additivity of coherent information under two-way processing. These results establish Riemannian manifold optimization as a practical tool for probing the physical limits of future quantum networks.
\end{abstract}

\maketitle

% \textit{Introduction}---Quantum entanglement constitutes a cornerstone of quantum information science \cite{horodecki_quantum_2009}, serving as the indispensable resource for critical protocols ranging from quantum teleportation \cite{peres_optimal_1991,bennett_purification_1996,murao_quantum_1999,murao_quantum-information_2000,studzinski_port-based_2017,mozrzymas_optimal_2018,zomorodi-moghadam_optimizing_2018,daei_improving_2021,christandl_asymptotic_2021,qiu_quantum_2022,strelchuk_minimal_2023,kim_asymptotic_2024,wills_efficient_2024}, superdense coding \cite{bennett_communication_1992}, to quantum cryptography \cite{ekert_quantum_1991,pirandola_advances_2020,portmann_security_2022}. As the field matures toward scalable implementations, the focus has shifted to distributed quantum information processing, where distant local processors form the nodes of a global quantum network \cite{beals_efficient_2013,zhao_practical_2021,caleffi_distributed_2024,barral_review_2025}. In this distant-lab paradigm, the class of local operations and classical communication (LOCC) \cite{peres_optimal_1991,chitambar_everything_2014} is not merely a theoretical construct but the fundamental operational restriction. Consequently, characterizing the limits of entanglement manipulation under LOCC is a central problem in network-based quantum computing.
\textit{Introduction}---Quantum entanglement is a cornerstone of quantum information science \cite{horodecki_quantum_2009}, underpinning essential protocols such as quantum teleportation \cite{bennett_purification_1996}, superdense coding \cite{bennett_communication_1992}, and quantum cryptography \cite{ekert_quantum_1991}. As the field advances toward scalable implementations, the focus has shifted to distributed quantum information processing, where spatially separated nodes form a global quantum network \cite{beals_efficient_2013,zhao_practical_2021}. In this setting, local operations and classical communication (LOCC) \cite{chitambar_everything_2014} is the fundamental operational restriction. Characterizing the limits of entanglement manipulation under LOCC is therefore a central problem in network-based quantum computing.

Despite its operational universality, the mathematical structure of LOCC remains notoriously elusive \cite{chitambar_everything_2014}. The set of LOCC operations is not topologically closed and exhibits a complex geometric structure, rendering the design and optimization of such protocols intractable \cite{chitambar_round_2017}. While LOCC-assisted key tasks like entanglement distillation \cite{bennett_purification_1996,deutsch_quantum_1996,murao_multiparticle_1998,dur_entanglement_2007,pan_experimental_2003,devetak_distillation_2005,czechlewski_distillation_2012,wang_improved_2016,de_bone_ghz_2023,miguel-ramiro_quantum_2023,du_advantage_2025,rozgonyi_practical_2025,zhu_geometric_2025}, entanglement-assisted teleportation \cite{peres_optimal_1991,bennett_purification_1996,murao_quantum_1999,murao_quantum-information_2000,studzinski_port-based_2017,mozrzymas_optimal_2018,zomorodi-moghadam_optimizing_2018,daei_improving_2021,christandl_asymptotic_2021,qiu_quantum_2022,strelchuk_minimal_2023,kim_asymptotic_2024,wills_efficient_2024}, state discrimination \cite{bennett_quantum_1999,walgate_local_2000,calsamiglia_local_2010,bandyopadhyay_more_2011,chitambar_when_2013,chitambar_revisiting_2013,childs_framework_2013,bandyopadhyay_entanglement_2016,halder_strong_2019,leung_locc_2021,zhu_entanglement_2025}, state redistribution \cite{horodecki_partial_2005,horodecki_quantum_2006,devetak_exact_2008,berta_single-shot_2009,bjelakovic_universal_2013,berta_smooth_2016,streltsov_entanglement_2016,yamasaki_quantum_2019,streltsov_quantum_2020}, and channel simulation \cite{bennett_quantum_2014,wilde_entanglement_2018,fang_quantum_2019,wang_exact_2023,cao_quantum_2024,cao_channel_2024} have been extensively studied, our understanding remains fragmented. Fundamental questions—such as the optimal strategies for finite-copy entanglement distillation and mixed state discrimination—remain open, primarily because navigating the non-convex landscape of LOCC protocols is computationally prohibitive.

To circumvent these structural complexities, a common approach to study the power and limit of LOCC protocols is to relax the LOCC to a larger set, e.g., separable operations \cite{chitambar_increasing_2012,gheorghiu_entanglement_2007,chitambar_nonlocal_2009,cohen_necessary_2014,chen_separable_2014,mancinska_separable_2013,hebenstreit_maximally_2016,duan_distinguishability_2009}, positive partial transpose (PPT) maps \cite{wang_improved_2016,fang_quantum_2019,wang_cost_2020,wang_exact_2023,zhu_entanglement_2025}, and $k$-extendible channels \cite{pankowski_entanglement_2011,kaur_extendibility_2019,singh_unextendible_2025}. These relaxations facilitate theoretical analysis and, particularly in the cases of PPT and $k$-extendibility, allow for the efficient computation of performance bounds via semi-definite programming (SDP) \cite{berta_semidefinite_2022}. However, these methods suffer from an ``operational gap'': they yield outer bounds rather than physically implementable protocols. Conversely, recent heuristic methods that utilize parameterized quantum circuits (PQCs) \cite{zhao_practical_2021,su_design_2023,rozgonyi_training_2024} seek explicit protocols but are often plagued by barren plateaus \cite{mcclean_barren_2018} and limited ansatz expressibility. Thus, the community lacks a unified framework capable of efficiently exploring the full space of fixed-round LOCC to discover implementable protocols and establish tight achievable bounds, without being confined by restricted circuit ansatzes or loose relaxations.

In this Letter, we bridge this gap by mapping fixed-round LOCC protocols directly onto the product Stiefel manifold. By characterizing the recursive structure of quantum instruments, we convert a constrained optimization over LOCC strategies into an unconstrained Riemannian optimization~\cite{boumal_introduction_2023,rapcsak_minimization_2002,tagare_notes_2011,li_efficient_2020,li_quantum_2025}. This geometric reformulation enables the direct discovery of LOCC protocols—including instruments with post-selection (IPS) and channel-measurement schemes (CMPS)—by exploring the full operational landscape of finite-round LOCC without recourse to relaxations or restricted circuit ansatz.
% Importantly, it does not rely on loose relaxations or restricted circuit ansatzes.

We demonstrate this framework through entanglement distillation. Our method quantifies the operational advantage of adaptive communication rounds, yields achievable fidelities that match PPT upper bounds to within numerical precision, and provides numerical evidence for super-additivity of coherent information under two-way processing. These results establish Riemannian optimization on physically structured manifolds as a practical route for designing protocols and probing the limits of future quantum networks.

\textit{Geometric Embedding of LOCC Physical Constraints}---The operational dynamics of fixed-round LOCC are described by quantum instruments—collections of completely positive (CP) maps that capture both probabilistic classical outcomes and post-measurement state updates. An instrument $\mathfrak{J} = \{\mathcal{E}_j\}_{j=1}^S$ acts on a local system via a family of completely positive (CP) maps. Probability conservation dictates that the sum of these maps is trace-preserving (TP), implying the physical constraint $\sum_{j=1}^S \sum_{i=1}^{T_j} K_{j,i}^\dagger K_{j,i} = \mathbb{I}_d$, where $K_{j,i}$ are the Kraus operators. By vertically stacking all Kraus operators into a single matrix $\mathbf{K}$, the set of valid local instruments is naturally parameterized by the Stiefel manifold
\begin{equation}
    \mathrm{St}(D, d) := \{\mathbf{K} \in \mathbb{C}^{D \times d} \vert \mathbf{K}^\dagger \mathbf{K} = \mathbb{I}_d\}
\end{equation}
where $D = d \sum_{j=1}^S T_j$. This step embeds the physical CPTP constraints of quantum instruments into a continuous Riemannian geometric structure.

In each round of an LOCC protocol, a designated party $X$ performs a local instrument $\mathfrak{J}^{(X)} = \{\mathcal{E}_j^{(X)}\}_{j=1}^S$ on $\mathcal{H}_X$, obtaining outcome $j \in [S]$, and broadcasts $j$ to all other parties; upon receiving $j$, each party $J \ne X$ applies a conditional CPTP map $\mathcal{T}_j^{(J)}$ on $\mathcal{H}_J$. This round of operation is captured by the formalism of a \textit{one-way local instrument} \cite{chitambar_everything_2014}, $\mathfrak{J}_{\mathrm{owl}}^{(X)} = (\mathcal{A}_j)_{j=1}^S$, where
\begin{equation}
    \mathcal{A}_j = \left(\bigotimes_{J\ne X} \mathcal{T}_j^{(J)}\right) \otimes \mathcal{E}_j^{(X)},
\end{equation}
with $\mathcal{E}_j^{(X)}$ a CP map on $\mathcal{H}_X$ encoding the measurement and $\mathcal{T}_j^{(J)}$ the conditional CPTP response of party $J$. Since each $\mathfrak{J}^{(X)}$ and $\mathcal{T}_j^{(J)}$ individually embeds onto a Stiefel manifold, their joint feasible set forms a product manifold:

\begin{proposition}[One-way local instruments]
Given a one-way local instrument $\mathfrak{J}_{\mathrm{owl}}^{(X)}$ with Kraus rank $T_j^{(X)}$ for $\mathcal{E}_j^{(X)}$ and Kraus rank $T_j^{(J)}$ for $\mathcal{T}_j^{(J)}$, the feasible set of $\mathfrak{J}_{\mathrm{owl}}^{(X)}$ is the product Stiefel manifold
\begin{align}
    \mathcal{M}_{\mathrm{owl}}^{(X)} = \left[\bigtimes_{j\in[S], J\ne X}\mathrm{St}(D_j^{(J)}, d_J)\right] \times \mathrm{St}(D^{(X)}, d_X),
\end{align}
where $d_X = \dim\mathcal{H}_X$, $d_J = \dim\mathcal{H}_J$, $D_j^{(J)} = T_j^{(J)} d_J$, and $D^{(X)} = d_X \sum_{j=1}^S T_j^{(X)}$.
\end{proposition}

\begin{figure}
    \centering
    \includegraphics[width=0.95\linewidth]{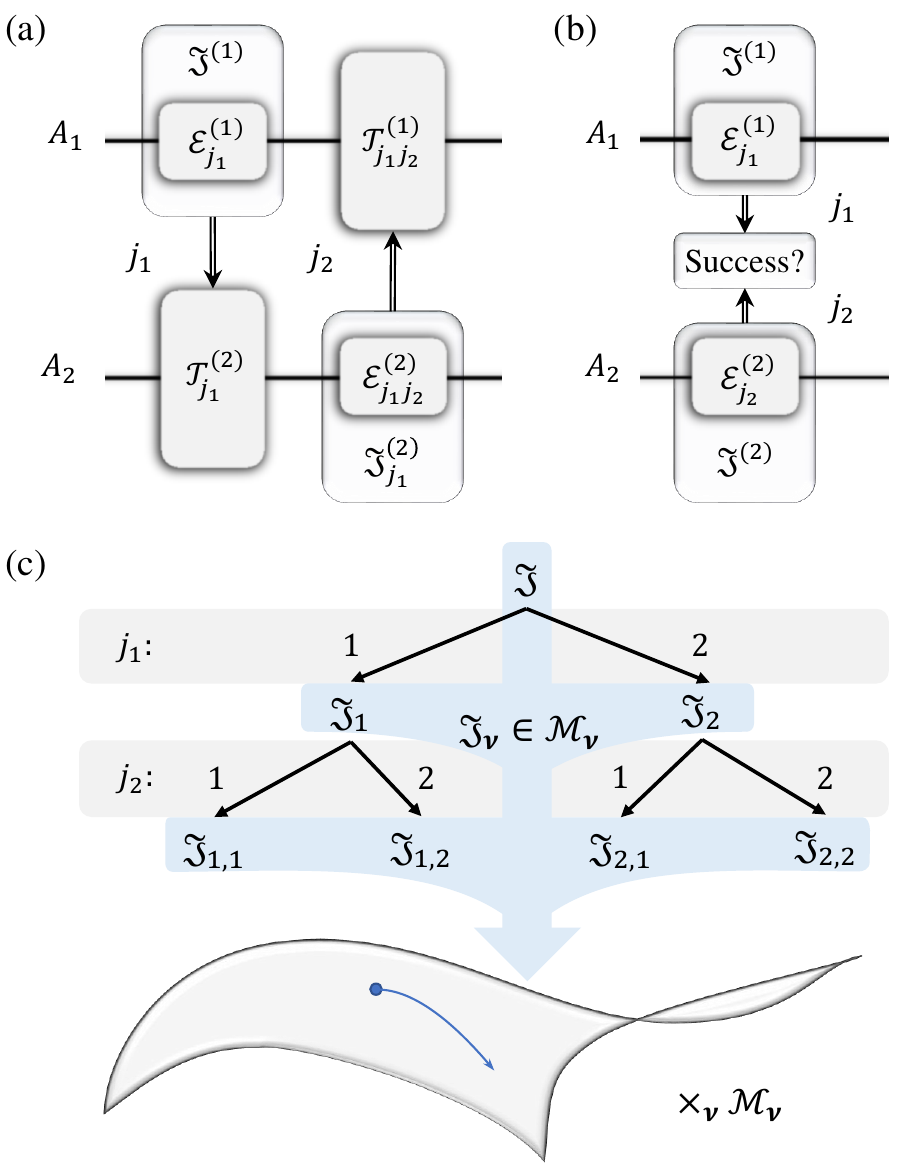}
    \caption{Operational structures of LOCC. (a) A two-round bipartite example. Round 1 begins with $A_1$ applying a local instrument and broadcasting outcome $j_1$, which determines $A_2$'s local CPTP map. In Round 2, $A_2$ performs a history-dependent instrument yielding $j_2$, triggering $A_1$'s final CPTP update. (b) The IPS scheme. Parties independently apply local instruments and broadcast outcomes, with the protocol declared successful only upon obtaining a specific outcome sequence. (c) The general LOCC decision tree (for $S=2$), where each node $\mathcal{J}_{\bm{\nu}}$ on the product Stiefel manifold $\cM_{\bm{\nu}}$ represents the one-way local instrument conditioned on history $\bm{\nu}$. Optimizing the LOCC protocol corresponds to the optimization on the product Stiefel manifold $\times_{\bm{\nu}}\cM_{\bm{\nu}}$}
    \label{fig:LOCC_structure}
\end{figure}

An $r$-round LOCC protocol $\locc_r$ is then constructed by composing one-way local instruments across $r$ rounds, where each round's instrument is conditioned on the cumulative history of prior outcomes. This adaptive structure forms a decision tree [Fig.~\ref{fig:LOCC_structure}(c)], where each node $\bm{j} = [j_1, \dots, j_{r-1}]$ specifies a history path and the corresponding one-way local instrument $\mathfrak{J}_{\mathrm{owl},\bm{j}}^{(X_r)}$ to be executed at that node. Since each node's feasible set is itself a product Stiefel manifold $\mathcal{M}_{\mathrm{owl},\bm{j}}^{(X_r)}$ (Proposition 1), the full protocol's feasible set is the Cartesian product over all nodes across the tree:

\begin{proposition}[Fixed-round LOCC]
Given an $r$-round LOCC protocol $\locc_r$ where party $X_k$ is the designated measuring party in round $k$ for $k = 1, \dots, r$, the feasible set of $\locc_r$ is the recursive product Stiefel manifold
\begin{align}
    \mathcal{M}_{\locc_r}^{(X_r,\dots,X_1)} = \left[\bigtimes_{\bm{j}}\mathcal{M}_{\mathrm{owl}, \bm{j}}^{(X_r)}\right] \times \mathcal{M}_{\locc_{r-1}}^{(X_{r-1},\dots,X_1)},
\end{align}
where $\mathcal{M}_{\mathrm{owl}, \bm{j}}^{(X_r)}$ is the product Stiefel manifold of the one-way local instrument at node $\bm{j}$ (Proposition~1), with base case $\mathcal{M}_{\locc_0} = \{\mathrm{id}\}$ denoting the trivial protocol consisting solely of the identity channel.
\end{proposition}

An important special case of $\locc_N$ is the \textit{Instruments with Post-Selection} (IPS) scheme [Fig.~\ref{fig:LOCC_structure}(b)]. In many practical quantum network settings, real-time adaptive classical communication between rounds is unavailable or prohibitively costly; parties are instead constrained to perform their local measurements simultaneously and independently, declaring success only if the joint outcome matches a pre-specified sequence. This physical restriction is precisely captured by an id-local instrument $\mathfrak{J}_{\mathrm{idl}} = \bigotimes_{X \in [N]} \mathfrak{J}^{(X)}$, where each party $X$ performs an independent instrument $\mathfrak{J}^{(X)} = \{\mathcal{E}_j^{(X)}\}_{j=1}^{S_X}$ with Kraus rank $T_j^{(X)}$ without conditioning on any other party's outcome. As a subclass of $\locc_N$ without inter-round communication, IPS serves as a natural baseline for quantifying the operational advantage conferred by adaptive rounds. Since the instruments are independent and unconditioned, the IPS feasible set is simply the product of individual Stiefel manifolds:

\begin{proposition}[IPS structure]
Given an IPS protocol where each party $X \in [N]$ independently applies an instrument $\mathfrak{J}^{(X)}$ with Kraus rank $T_j^{(X)}$ per outcome $j \in [S_X]$, the feasible set of IPS is the product Stiefel manifold
\begin{align}
    \mathcal{M}_{\mathrm{IPS}} = \mathrm{St}(D^{(1)}, d_1) \times \dots \times \mathrm{St}(D^{(N)}, d_N),
\end{align}
where $D^{(X)} = d_X \sum_{j=1}^{S_X} T_j^{(X)}$ is the ambient dimension for party $X$'s instrument.
\end{proposition}

A further restriction is the \textit{Channel-Measurement Post-Selection} (CMPS) scheme, a subclass of IPS that reflects a ubiquitous experimental constraint: the measurement apparatus $\mathfrak{M}^{(X)}$ (e.g., a Bell-state measurement station or a fixed-basis detector) is predetermined by the hardware, while each party retains freedom only in designing a preprocessing CPTP channel $\mathcal{T}^{(X)}$. This decomposition captures the common experimental paradigm of local preprocessing followed by a fixed readout, and generalizes parameterized quantum circuit (PQC) approaches such as LOCCNet \cite{zhao_practical_2021}, in the sense that $\mathrm{LOCCNet} \subseteq \mathrm{CMPS}$. Fixing $\mathfrak{M}^{(X)}$ constrains the feasible set to the Kraus operators of $\mathcal{T}^{(X)}$ alone, embedding $\mathcal{M}_{\mathrm{CMPS}}$ as a submanifold of $\mathcal{M}_{\mathrm{IPS}}$:

\begin{proposition}[CMPS structure]
Given a CMPS protocol where each party $X$ applies a fixed measurement instrument $\mathfrak{M}^{(X)}$ composed with a CPTP channel $\mathcal{T}^{(X)}$ of Kraus rank $T^{(X)}$, the feasible set of CMPS is the product Stiefel manifold
\begin{align}
    \mathcal{M}_{\mathrm{CMPS}} = \mathrm{St}(D'^{(1)}, d_1) \times \dots \times \mathrm{St}(D'^{(N)}, d_N),
\end{align}
where $D'^{(X)} = T^{(X)} d_X$.
\end{proposition}

Taken together, propositions 1--4 establish a unified geometric framework: the hierarchy $\mathcal{M}_{\locc_r} \supset \mathcal{M}_{\mathrm{IPS}} \supset \mathcal{M}_{\mathrm{CMPS}}$ reflects the increasing operational restrictions on the protocol, with each level admitting an explicit product Stiefel manifold parameterization. This transforms the intractable constrained optimization of fixed-round LOCC into an efficient, unconstrained Riemannian optimization problem, enabling the direct discovery of implementable, locally optimal protocols.

\begin{figure*}[t]
    \centering
    % 第一行两张图
    \includegraphics[width=0.95\textwidth]{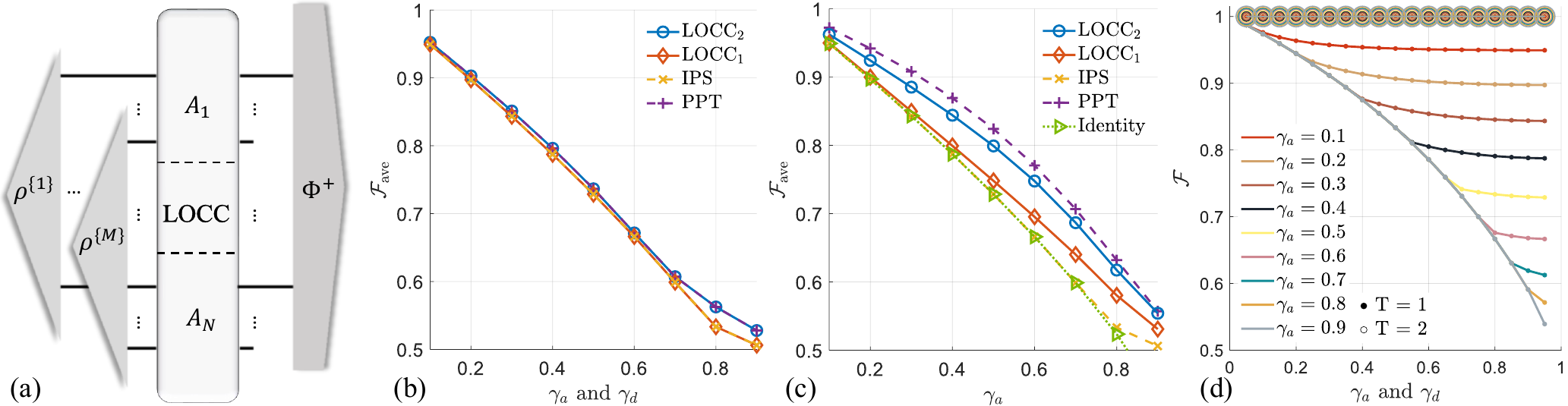}
    \caption{Achievable benchmarks for LOCC-assisted entanglement distillation. 
{(a)} Schematic of $N$-partite entanglement distillation from $M$ states. Here, input states are MESs after given noise channels. 
{(b, c)} Optimization of the average distillation fidelity $\cF_{\rm ave}$ comparing IPS, $\locc_1$, and $\locc_2$ schemes against the PPT upper bound, where $\locc_2$ outperforms $\locc_1$, providing numerical evidence of the round advantage.
Panel (b) depicts the non-\textit{i.i.d.} regime (copy 1: amplitude damping parametrized by $\gamma_a$; copy 2: depolarizing parametrized by $\gamma_d$), where $\locc_2$ matches the PPT bound within numerical precision. 
Panel (c) shows the results for \textit{i.i.d.} amplitude damping noise. 
{(d)} Probabilistic distillation performance via the CMPS scheme for the non-\textit{i.i.d.} setting in (b). 
Notably, with Kraus rank $T=2$, the fidelity approaches unity, suggesting that near-perfect entanglement can be distilled by trading off success probability.}
    \label{fig:dstl_fid_res}
\end{figure*}

% \section{Achievable Benchmarks for Entanglement Distillation}
\textit{Entanglement Distillation Fidelity}---Having embedded fixed-round LOCC protocols into a product Stiefel manifold, we now connect the theoretical construction to an experimentally meaningful benchmark: the fidelity of finite-copy entanglement distillation. Entanglement distillation converts multiple copies of noisy entangled states into fewer copies of high-fidelity maximally entangled states (MES) \cite{peres_optimal_1991,bennett_purification_1996,murao_quantum_1999,murao_quantum-information_2000,studzinski_port-based_2017,mozrzymas_optimal_2018,zomorodi-moghadam_optimizing_2018,daei_improving_2021,christandl_asymptotic_2021,qiu_quantum_2022,strelchuk_minimal_2023,kim_asymptotic_2024,wills_efficient_2024,bennett_communication_1992,ekert_quantum_1991,pirandola_advances_2020,portmann_security_2022}.  
The distilled entanglement then serves as a resource for teleportation, superdense coding, and quantum key distribution. In practice, only a finite number of noisy copies are available, rendering asymptotic rate formulas inapplicable and making the finite-copy protocol structure the determining factor of performance. 
While outer bounds derived from relaxations (e.g., PPT or $k$-extendibility) are well-studied~\cite{rains_bound_1999,fang_non-asymptotic_2019,rozpedek_optimizing_2018,kaur_extendibility_2019}, they fail to yield implementable protocols. Our geometric framework bridges this gap by directly optimizing fixed-round LOCC operations on the product Stiefel manifold, enabling the discovery of implementable protocols and tight achievable bounds. Notably, unlike conventional studies focused on symmetric \textit{i.i.d.} sources, our framework naturally allows us to benchmark protocols for both \textit{i.i.d.} and non-\textit{i.i.d.} inputs.

\begin{proposition}
For $N$ parties sharing $M$ copies of noisy states $\rho$, an LOCC protocol $\mathcal{N}{\mathbf{K}}$ parameterized by $\mathbf{K} \in \mathcal{M}$, and a success set $\mathcal{S}$ of outcome sequences, the optimal distillation fidelity is
\begin{align}
    \cF^*(\mathcal{S}) = \max_{\mathbf{K} \in \mathcal{M}} \frac{\sum_{\bm{j}\in\mathcal{S}} \bra{\Phi^+} \rho_{\bm{j}}(\mathbf{K}) \ket{\Phi^+}}{\sum_{\bm{j}\in\mathcal{S}} \Tr[\rho_{\bm{j}}(\mathbf{K})]},
\end{align}
where $\rho_{\bm{j}}(\mathbf{K}) = \mathcal{N}{\mathbf{K}}(\rho^{\otimes M})\vert{\bm{j}}$ is the sub-normalized output conditioned on outcome $\bm{j}$. If $\mathcal{S}$ is the set of all outcomes, the denominator equals $1$ by trace preservation, recovering the average fidelity $\cF_{\rm ave}$. If $\mathcal{S}$ is a designated success set, the denominator is the success probability. In either case, this is an unconstrained Riemannian optimization over $\mathcal{M} \in \{\mathcal{M}{\locc_r}, \mathcal{M}{\mathrm{IPS}}, \mathcal{M}_{\mathrm{CMPS}}\}$.
\end{proposition}

\begin{figure}
    \centering
    \includegraphics[width=0.85\linewidth]{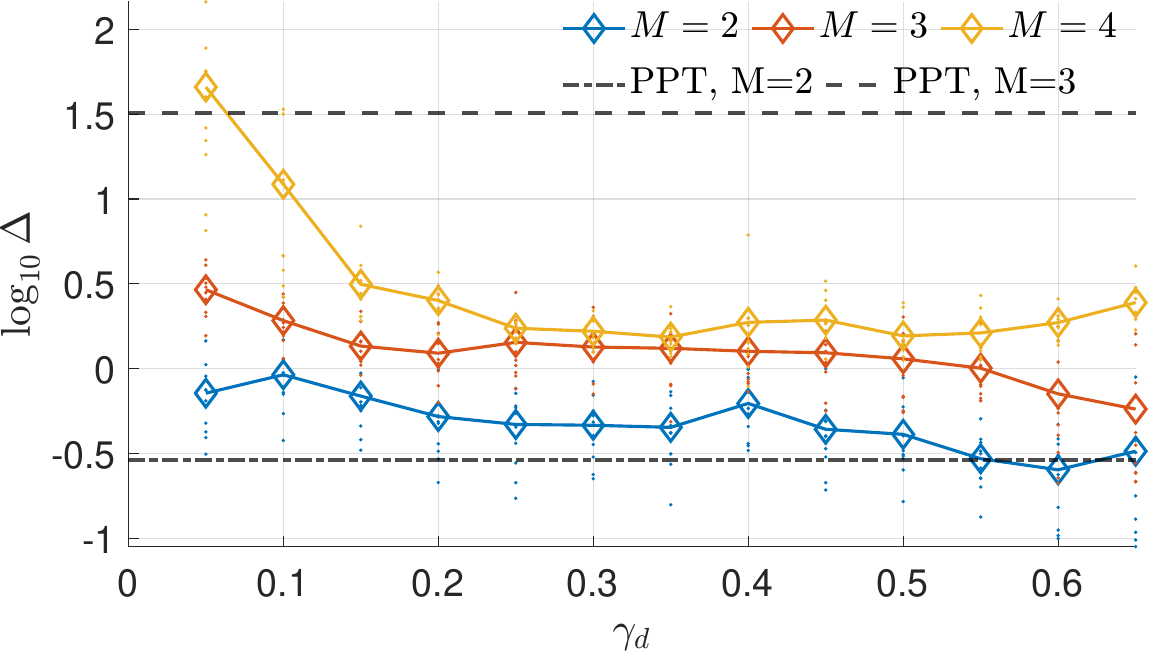}
    \caption{Logarithmic running time for distilling $M$ copies of depolarized states parameterized by $\gamma_d$ (10 trials per setting). 
Solid lines (dots) denote the average (individual) times for CMPS, while dashed lines indicate the minimal times for PPT via SDP. 
We do not present PPT results for $M=4$ as the computation could not be completed within a reasonable timeframe. This highlights the efficiency of our proposed framework.}
    \label{fig:dstl_time}
\end{figure}

To characterize the performance of distillation protocols, we consider three noise models: the depolarizing channel $\cN_{\mathrm{depo.}}$, the amplitude-damping channel $\cN_{\mathrm{a.d.}}$, and the dephasing channel $\cN_{\mathrm{deph.}}$, parameterized by $\gamma_d, \gamma_a, \gamma_p \in [0,1]$ respectively:
\begin{align}
    &\cN_{\mathrm{depo.}}(\gamma_d, \Phi^+) = (1-\gamma_d)\Phi^+ + \gamma_d \idop/d,\\
    &\cN_{\mathrm{a.d.}}(\gamma_a, \Phi^+) \!=\! K_0 \Phi^+ K_0^\dagger + \gamma_a \sum_{i=1}^{d-1} \ketbra{i\!-\!1}{i}\Phi^+\ketbra{i}{i\!-\!1},\\
    &\cN_{\mathrm{deph.}}(\gamma_p, \Phi^+) = \gamma_p \hat{\Phi}^+ + (1-\gamma_p)\Phi^+,
\end{align}
where $K_0 = \ketbra{0}{0} + \sum_{i=1}^{d-1}\sqrt{1-\gamma_a}\ketbra{i}{i}$ and $\hat{\Phi}^+$ denotes $\Phi^+$ with off-diagonal elements set to zero.

We first consider deterministic distillation ($\mathcal{S}$ = all outcomes), where the protocol must deliver entanglement regardless of measurement outcomes—a requirement for network applications that demand guaranteed entanglement supply. Here, Proposition~5 reduces to the average fidelity $\cF_{\rm ave}$. We benchmark three protocol classes—IPS, $\locc_1$, and $\locc_2$—on $N$-party systems sharing $M$ copies of noisy Bell states [Fig.~\ref{fig:dstl_fid_res}(a)]. Results are shown in Fig.~\ref{fig:dstl_fid_res}(b,c). In the non-i.i.d. setting (copy 1: amplitude damping; copy 2: depolarizing), we find $\cF_{\rm ave}(\locc_2) > \cF_{\rm ave}(\locc_1) \approx \cF_{\rm ave}(\text{IPS})$. Notably, the optimized $\locc_2$ fidelity matches the PPT upper bound to within numerical precision \cite{rains_bound_1999,fang_non-asymptotic_2019}, suggesting optimality for these instances. The gap between $\locc_2$ and $\locc_1$ provides numerical evidence of a round advantage: the second round of adaptive communication enables error-correction strategies inaccessible to single-round protocols.
% Conversely, for systems subject to strong depolarizing or dephasing noise, the achievable fidelity collapses to that of the identity operation, suggesting that the investigated schemes yield no distillation gain in these input settings (see Supplemental Material [SM]).

We then consider the probabilistic case ($\mathcal{S}$ = a designated success set), where post-selection is permitted, and the protocol declares success only upon obtaining a favorable outcome. This setting reflects the experimental paradigm of heralded entanglement generation, where the cost of a failed attempt can be amortized through repetition, making it worthwhile to trade success probability for near-perfect fidelity. Applying Proposition~5 with the CMPS scheme, we numerically optimize the post-selected fidelity for a specified outcome sequence. In the non-\textit{i.i.d.} setting [Fig.~\ref{fig:dstl_fid_res}(d)], the results suggest that a Kraus rank of $T=2$ can already yield fidelities approaching unity, motivating the conjecture that Kraus-rank-2 CMPS protocols may suffice for complete distillation in this setting. We further apply this optimization to multi-copy distillation, obtaining protocols for bipartite distillation with up to $M=6$ copies and tripartite distillation with up to $M=4$ copies. This scalability is rooted in the computational efficiency of the geometric framework: as demonstrated in Fig.~\ref{fig:dstl_time}, it produces achievable protocols for the distillation problem with an efficiency orders of magnitude higher than SDP-based PPT relaxations. {Rather than merely computing unachievable outer bounds, this efficiency empowers the direct discovery of complex, multi-copy LOCC protocols, unlocking a critical pathway for scaling up quantum network architectures.}

\textit{Two-Way Distillation Rate}---
Beyond the finite-copy regime, characterizing the asymptotic limit of the two-way distillable entanglement, $D_\leftrightarrow(\rho)$, is a fundamental pursuit in distributed quantum information processing. This ultimate limit is governed by the regularized formula~\cite{devetak_distillation_2005}:
{
\begin{equation}
D_\leftrightarrow(\rho) = \lim_{n \rightarrow \infty} \frac{1}{n} D_{\leftrightarrow}^{(1)}(\rho^{\otimes n}),
\end{equation}
% where $D_{\leftrightarrow}^{(1)}(\rho^{\otimes n})$ denotes the one-shot two-way distillable entanglement for an $n$-copy input, representing the maximum rate extractable via general two-way LOCC instruments. According to the hashing bound principle, this quantity can be lower-bounded by the maximum block-length coherent information:
where
\begin{equation}\label{eq:two_way_coh_info}
D_{\leftrightarrow}^{(1)}(\rho^{\otimes n}) \coloneqq \max_{\Lambda \in \mathrm{LOCC}_2} I(A'\rangle B')_{\Lambda(\rho^{\otimes n})}.
\end{equation}
}
Evaluating this limit is difficult because the regularization $n \to \infty$ cannot be directly computed, and the non-convex structure of the $\locc_2$ set makes analytical characterization challenging.

\begin{remark}
Given an $n$-copy input state $\rho^{\otimes n}$ and a two-way LOCC protocol $\mathcal{N}_{\mathbf{K}}$ parameterized by $\mathbf{K} \in \mathcal{M}_{\locc_2}$, the normalized coherent information $\frac{1}{n} I(A'\rangle B')_{\mathcal{N}_{\mathbf{K}}(\rho^{\otimes n})}$ provides an achievable lower bound on the two-way distillable entanglement. Optimizing this quantity over $\mathcal{M}_{\locc_2}$ gives a variational lower-bound problem, for which Riemannian optimization returns numerically optimized achievable values.
\end{remark}

We apply this to the Choi state $\rho_{AB} = \mathcal{I} \otimes \mathcal{N}_{\mathrm{g.a.d.}}(\Phi_{AB}^+)$ of the generalized amplitude damping channel (GADC) parameterized by damping $\gamma_a \in [0,1]$ and thermal noise $\gamma_n \in [0,1]$ (reducing to standard amplitude damping when $\gamma_n=0$), whose Kraus operators are
\begin{equation}
\begin{aligned}
 &K_1 = \sqrt{1-\gamma_n}(\ketbra{0}{0} + \sqrt{1-\gamma_a}\ketbra{1}{1}),\\
&K_2 = \sqrt{\gamma_n}(\sqrt{1-\gamma_a}\ketbra{0}{0} + \ketbra{1}{1}),\\
  &K_3 = \sqrt{\gamma_a(1-\gamma_n)}\ketbra{0}{1}, ~K_4 = \sqrt{\gamma_a\gamma_n}\ketbra{1}{0}.
\end{aligned}
\end{equation}

\begin{figure}
    \centering
    \includegraphics[width=0.85\linewidth]{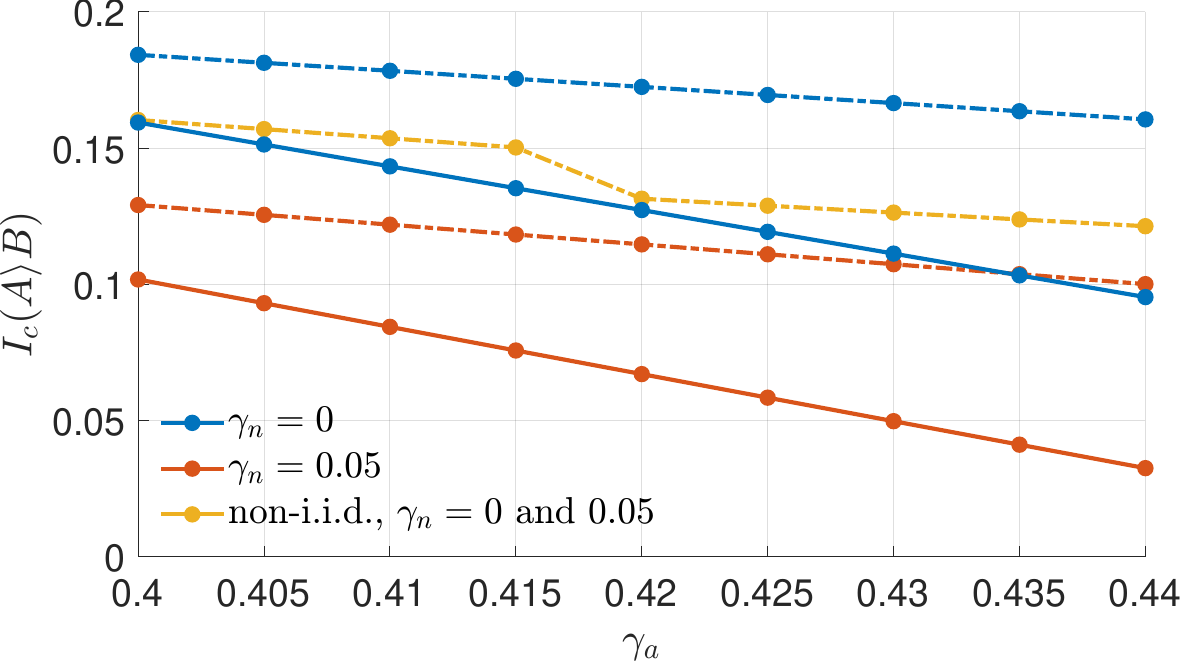}
    \caption{Coherent information optimization for GADC channels.
Solid and dashed lines represent the single-copy Hashing bound and the optimized two-copy results, respectively.
Non-\textit{i.i.d.} inputs utilize $\gamma_n=0$ and $0.05$.
The two-copy results lie above the Hashing bound in the tested regimes, suggesting improved achievable lower bounds for distillable entanglement and providing numerical evidence for the super-additivity of coherent information.
}
    \label{fig:dstl_ent}
\end{figure}

We numerically optimize the normalized coherent information over $\mathcal{M}_{\locc_2}$ to obtain finite-block achievable lower bounds. As a natural baseline, the hashing bound is given by the coherent information $I(A\rangle B)_{\rho}$ of the unassisted state. When optimizing a single copy ($n=1$) over the $\locc_2$ manifold, our numerical results do not show an improvement over the hashing bound, suggesting that local preprocessing may offer limited advantage for this specific noise model in the single-shot regime. In contrast, optimizing the joint processing of two copies ($n=2$) on the product Stiefel manifold yields a per-copy coherent information that exceeds the single-copy baseline in the tested parameter regimes (see Fig.~\ref{fig:dstl_ent}). This observed gain isolates the advantage of joint processing within the optimized ansatz, providing numerical evidence for the super-additivity of coherent information under explicit $\locc_2$ constraints. Furthermore, the step-like behaviors observed in the non-\textit{i.i.d.} regimes suggest unexplored noise-induced transitions. Overall, these results highlight the power of the Riemannian optimization framework to uncover candidate super-additive effects and discover physically implementable protocols that improve known achievable lower bounds.

% \textit{Discussion}---\label{sec:discussion}
% We have introduced a product-Stiefel parameterization of fixed-round LOCC protocols, converting physical CPTP and communication constraints into an unconstrained Riemannian optimization problem. This provides a direct variational route to implementable LOCC protocols, complementing relaxation-based approaches that yield outer bounds but not explicit operations. The resulting geometric viewpoint may also guide the design of protocol ansatzes for larger multipartite networks.

% For entanglement distillation, the framework yields numerical evidence for adaptive-round advantages, protocols whose fidelities match PPT bounds within numerical precision, and finite-block coherent-information gains under two-way processing. The same construction can be combined with task-specific fidelity objectives beyond distillation, including one-shot state merging \cite{horodecki_partial_2005,horodecki_quantum_2006}; the corresponding formulation and numerical benchmarks are given in the Supplemental Material, Secs.~II and III C. These results suggest that Riemannian optimization over physically structured manifolds can serve as a practical route for exploring finite-resource quantum network protocols.

\textit{Discussion}---\label{sec:discussion}
We have introduced a product-Stiefel parameterization of fixed-round LOCC protocols, converting physical CPTP and communication constraints into an unconstrained Riemannian optimization problem. This provides a direct variational route to implementable LOCC protocols, complementing SDP-based relaxations that yield outer bounds but not explicit operations.

For entanglement distillation, the framework yields numerical evidence for adaptive-round advantages, protocols whose fidelities match PPT bounds to within numerical precision, and finite-block coherent-information gains under two-way processing—providing evidence for the super-additivity of coherent information under explicit LOCC constraints. Beyond distillation, the same geometric construction applies to any LOCC task with a fidelity-based objective. We have demonstrated this generality through one-shot state merging \cite{horodecki_partial_2005,horodecki_quantum_2006} (see Supplemental Material, Secs.~II and III~C), where the framework maps out achievable fidelity regions and reveals a numerical operational gap between CMPS and PPT bounds. These results establish that Riemannian optimization over physically structured manifolds is a practical and broadly applicable tool for exploring finite-resource quantum network protocols.

Several directions merit further investigation. First, while our numerical evidence for round advantage and super-additivity is compelling, analytical lower bounds on the number of rounds needed to saturate PPT bounds remain open. Second, extending the framework to incorporate experimental noise models and hardware-specific constraints could bridge the gap between theoretical protocol design and laboratory implementation. Third, the product-Stiefel structure may inspire new manifold-aware optimization algorithms with provable convergence guarantees for quantum information tasks.

\begin{acknowledgments}
We would like to thank Bin Gao, Xia Liu, Ben-Chi Zhao, and Xiao Shi for discussions. This work was partially supported by the National Key R\&D Program of China (Grant No. 2024YFB4504004); the National Natural Science Foundation of China (Grant No. 12447107); Guangdong Provincial Natural Science Foundation
(Grant. No.~2025A1515012834); the Guangdong Provincial Quantum Science Strategic Initiative (Grant No. GDZX2403008 and GDZX2403001); the Guangdong Provincial Key Lab of Integrated Communication, Sensing and Computation for the Ubiquitous Internet of Things (Grant No. 2023B1212010007); the Quantum Science Center of the Guangdong–Hong Kong–Macao Greater Bay Area; and the Education Bureau of Guangzhou Municipality.
\end{acknowledgments}

\bibliography{ref}% Produces the bibliography via BibTeX.

\clearpage
\appendix

% \section{SDP for PPT-assisted Distillation}
% The SDP for the PPT-assisted distillation to obtain the maximal average fidelity is given by
% \begin{align}
%     \max &\; \tr\Big[\rho_{AB}^T E_{AB}\Big],\\
% {\rm s.t.} &\;\; E,F\ge0\\
% &\;\; (1-d)F_{AB}^{T_{A}} \le E_{AB}^{T_B} \le (1+d)F_{AB}^{T_{A}},\\
% &\;\; E_{AB} + (d^2 - 1) \otimes F_{AB} = \idop_{AB}.
% \end{align}

% Given the success probability $p = \Tr[\rho_{AB}^{T}\Pi_{ABA'B'}]$, we have the SDP problem to obtain the maximal fidelity of a specified outcome sequence
% \begin{align}
%     \max &\; \tr\Big[\rho_{AB}^T E_{AB}\Big]/p,\\
% {\rm s.t.} &\;\; E_{AB}, F_{AB}\ge 0,\\
% &\Tr\{\rho_{AB}^{T}[E_{AB} + (d^2 - 1)F_{AB}]\} = p,\\
% &\;\; (1-d)F_{AB}^{T_{A}} \le E_{AB}^{T_B} \le (1+d)F_{AB}^{T_{A}},\\
% &\;\; E_{AB} + (d^2 - 1) \otimes F_{AB} \le \idop_{AB}.
% \end{align}

% \section{SDP for PPT-Assisted Merging}
% Let $C_{ABB'B''}$ denote the Choi state of the PPT operation. Then, the optimization problem for the average merging fidelity is given by
% \begin{align}
%     \max~~  & \Tr[C_{ABB'B''} \psi_{RAB}^{T_{AB}}\psi_{RB'B''}]\\
%     s.t.~~  & C_{ABB'B''}\ge 0, \label{const:cp}\\
%             & \Tr_{B'B''}[C_{ABB'B''}] = \idop_{AB}, \label{const:channel}\\
%             & C_{ABB'B''}^{T_{A}} \ge 0. \label{const:ppt}
% \end{align}

\end{document}

% \preprint{APS/123-QED}

\title{Supplemental Material: Optimizing LOCC Protocols on Product Stiefel Manifold}% 

\author{Ze-Tong Li}
 \affiliation{Thrust of Artificial Intelligence, Information Hub, The Hong Kong University of Science and Technology (Guangzhou), Guangdong 511453, China}
 \author{Cheng-Kai Zhu}
 \affiliation{Thrust of Artificial Intelligence, Information Hub, The Hong Kong University of Science and Technology (Guangzhou), Guangdong 511453, China}
\author{Xin Wang}%
 \email{felixxinwang@hkust-gz.edu.cn}
 \affiliation{Thrust of Artificial Intelligence, Information Hub, The Hong Kong University of Science and Technology (Guangzhou), Guangdong 511453, China}

\date{\today}% It is always \today, today,

\maketitle

\clearpage

\section{SDP for PPT-assisted Entanglement Distillation} \label{appendix:ppt_distillation}

\begin{figure}[h]
    \centering
    \includegraphics[width=.45\linewidth]{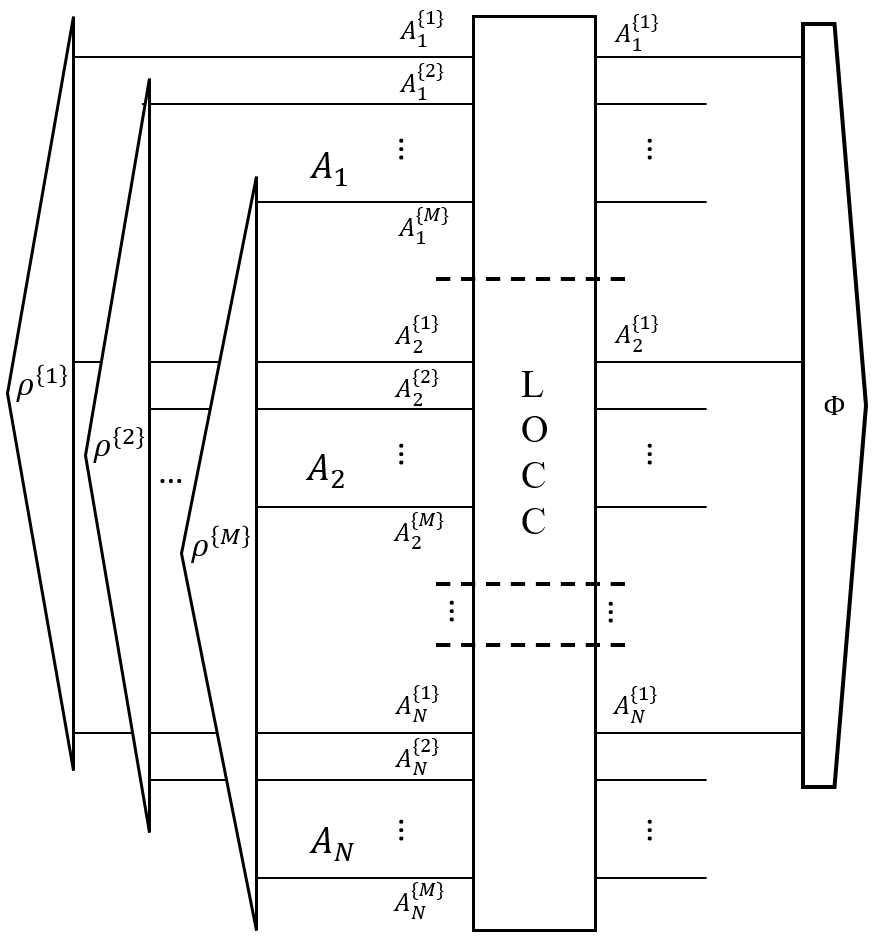}
    \caption{Schematic diagram of the $N$-agent $M$-copy LOCC-assisted entanglement distillation protocol.}
    \label{fig:ent_dstl_diagram}
\end{figure}

The schematic of entanglement distillation is illustrated in Fig.~\ref{fig:ent_dstl_diagram}. Let $\Pi_{ABA'B'}$ denote the Choi state corresponding to the PPT operation. Here, $A$ and $B$ represent the input systems for Alice and Bob, respectively, encompassing all $M$ copies of the input states $\rho_{AB}$. The systems $A'$ and $B'$ denote the $d$-dimensional output systems for Alice and Bob, respectively. The average fidelity of the protocol is given by
\begin{align}
    F = {\Tr[\rho_{AB}^{T}\Pi_{ABA'B'}\Phi_{A'B'}]},
\end{align}
where the identity operators on the ancillary systems are omitted. This leads to the following SDP optimization problem:
\begin{align}
    \max~~  & \Tr[\rho_{AB}^{T}\Pi_{ABA'B'}\Phi_{A'B'}],\\
    s.t.~~  & \Pi_{ABA'B'}\ge 0, \label{const:dstl_cp}\\
            & \Tr_{A'B'}[\Pi_{ABA'B'}] = \idop_{AB}, \label{const:dstl_channel}\\
            & \Pi_{ABA'B'}^{T_{AA'}} \ge 0, \label{const:dstl_ppt}
\end{align}
where constraints \eqref{const:dstl_cp} and \eqref{const:dstl_channel} ensure that $\Pi_{ABA'B'}$ represents the Choi state of a completely positive and trace-preserving (CPTP) map. Constraint \eqref{const:dstl_ppt} imposes the PPT condition, where $T_{AA'}$ denotes the partial transpose over the composite system $AA'$.

We observe that $\Pi_{ABA'B'}$ admits the following decomposition:
\begin{align}
    \Pi_{ABA'B'} = E_{AB} \otimes \Phi_{A'B'} + F_{AB} \otimes (\idop_{A'B'}-\Phi_{A'B'}).
\end{align}
Consequently, the CP constraint \eqref{const:dstl_cp} becomes
\begin{align}
    E_{AB} \otimes \Phi_{A'B'} + F_{AB} \otimes (\idop_{A'B'}-\Phi_{A'B'}) \ge 0.
\end{align}
Due to the orthogonality of $\Phi_{A'B'}$ and $(\idop_{A'B'}-\Phi_{A'B'})$, this inequality holds if and only if $E_{AB}, F_{AB}\ge0$. Similarly, the trace-preserving constraint \eqref{const:dstl_channel} transforms into
\begin{align}
    E_{AB} + (d^2 - 1)F_{AB} = \idop_{AB}.
\end{align}
Regarding the PPT constraint \eqref{const:dstl_ppt}, we derive:
\begin{align}
    &E_{AB}^{T_A} \otimes \Phi_{A'B'}^{T_{A'}} + F_{AB}^{T_A} \otimes (\idop_{A'B'}-\Phi_{A'B'}^{T_{A'}})\\
    =& E_{AB}^{T_A} \otimes \frac{P_+ - P_-}{d} + F_{AB}^{T_A} \otimes \frac{(d-1)P_+ + (d+1)P_-}{d}\\
    =&[E_{AB}^{T_{A}} + (d-1)F_{AB}^{T_{A}}]\frac{P_+}{d} + [-E_{AB}^{T_{A}} + (d+1)F_{AB}^{T_{A}}]\frac{P_-}{d}\\
    \ge& 0,
\end{align}
where $P_+$ and $P_-$ are the symmetric and anti-symmetric projection operators, respectively, with $\Phi_{A'B'}=(P_+ - P_-)/d$. The inequality holds if and only if
\begin{align}
    (1-d)F_{AB}^{T_{A}} \le E_{AB}^{T_A} \le (1+d)F_{AB}^{T_{A}}.
\end{align}

Combining these results, we arrive at the simplified SDP for PPT-assisted entanglement distillation
\begin{align}
    \max &\; \tr\Big[\rho_{AB}^T E_{AB}\Big],\\
{\rm s.t.} &\;\; E_{AB},F_{AB}\ge0\\
&\;\; (1-d)F_{AB}^{T_{A}} \le E_{AB}^{T_B} \le (1+d)F_{AB}^{T_{A}},\\
&\;\; E_{AB} + (d^2 - 1) F_{AB} = \idop_{AB}.
\end{align}

Next, we address the optimization of distillation fidelity where the protocol succeeds with probability $p$. The fidelity is defined as:
\begin{align}
    F = \frac{\Tr[\rho_{AB}^{T}\Pi_{ABA'B'}\Phi_{A'B'}]}{\Tr[\rho_{AB}^{T}\Pi_{ABA'B'}]}.
\end{align}
Fixing the success probability $p = \Tr[\rho_{AB}^{T}\Pi_{ABA'B'}]$, the optimization problem is formulated as:
\begin{align}
    \max~~  & \Tr[\rho_{AB}^{T}\Pi_{ABA'B'}\Phi_{A'B'}]/p,\\
    s.t.~~  & \Tr[\rho_{AB}^{T}\Pi_{ABA'B'}] = p,\\
            & \Pi_{ABA'B'}\ge 0,\\
            & \Tr_{A'B'}[\Pi_{ABA'B'}] \le \idop_{AB},\\
            & \Pi_{ABA'B'}^{T_{AA'}} \ge 0.
\end{align}
Applying the same reduction techniques, we obtain the corresponding simplified SDP:
\begin{align}
    \max &\; \tr\Big[\rho_{AB}^T E_{AB}\Big]/p,\\
{\rm s.t.} &\;\; E_{AB}, F_{AB}\ge 0,\\
&\Tr\{\rho_{AB}^{T}[E_{AB} + (d^2 - 1)F_{AB}]\} = p,\\
&\;\; (1-d)F_{AB}^{T_{A}} \le E_{AB}^{T_B} \le (1+d)F_{AB}^{T_{A}},\\
&\;\; E_{AB} + (d^2 - 1) F_{AB} \le \idop_{AB}.
\end{align}
\section{SDP for PPT-assisted State Merging} \label{appendix:ppt_state_merging}

\begin{figure}[h]
    \centering
    \includegraphics[width=.3\linewidth]{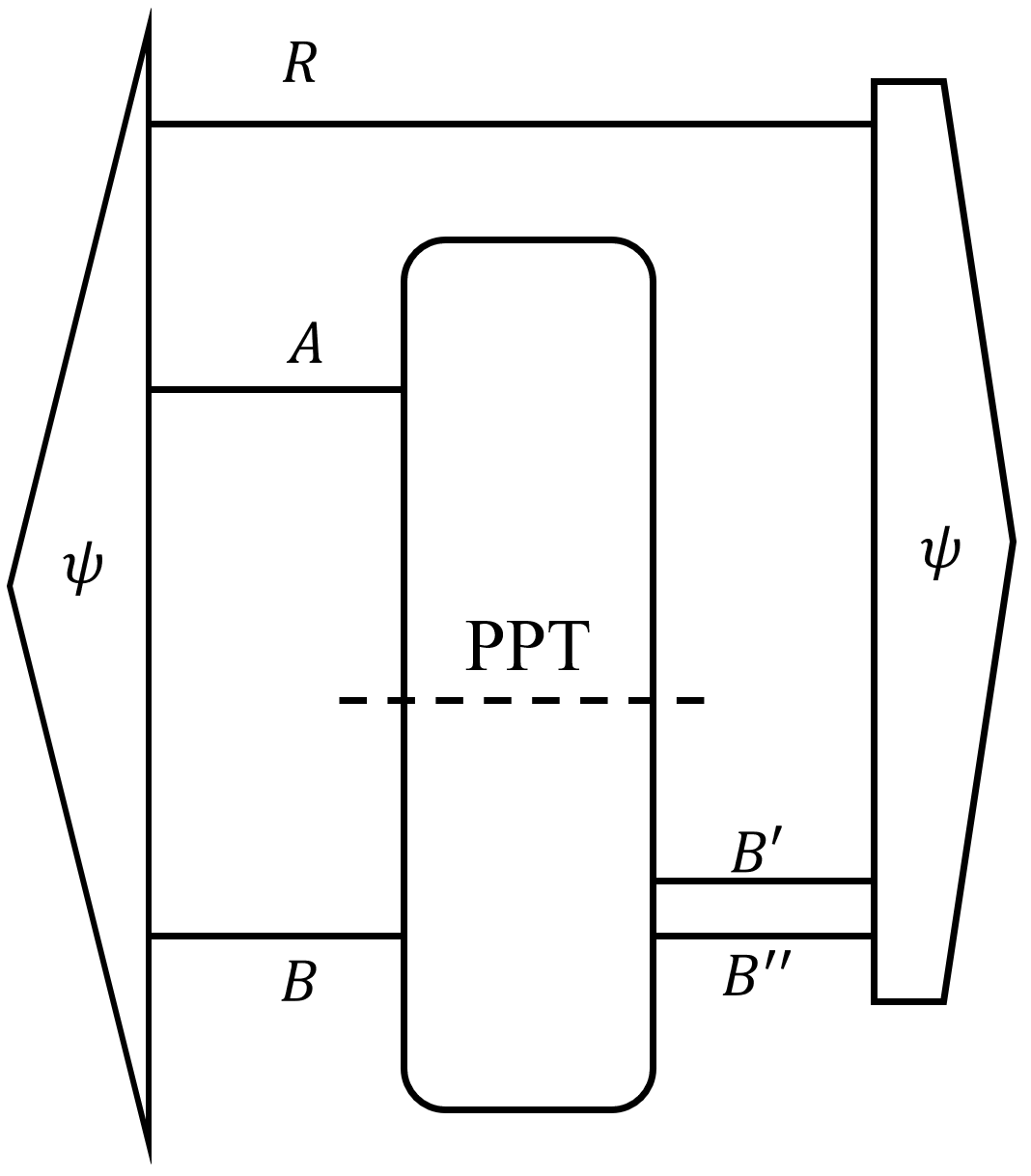}
    \caption{Schematic diagram of PPT-assisted state merging for the case $k=1$ and $m=1$.}
    \label{fig:state_merging_ppt}
\end{figure}

In this section, we present the SDP formulation for the PPT-assisted state merging task for the specific case where $k=1$ and $m=1$. The schematic representation of this protocol is depicted in Fig.~\ref{fig:state_merging_ppt}.

Let $C_{ABB'B''}$ denote the Choi state associated with the PPT operation. The optimization problem for maximizing the average merging fidelity is formulated as:
\begin{align}
    \max~~  & \Tr[C_{ABB'B''} \psi_{RAB}^{T_{AB}}\psi_{RB'B''}]\\
    s.t.~~  & C_{ABB'B''}\ge 0, \label{const:cp}\\
            & \Tr_{B'B''}[C_{ABB'B''}] = \idop_{AB}, \label{const:channel}\\
            & C_{ABB'B''}^{T_{A}} \ge 0, \label{const:ppt}
\end{align}
where constraints \eqref{const:cp} and \eqref{const:channel} ensure that $C_{ABB'B''}$ is a valid Choi state of a quantum channel. Constraint \eqref{const:ppt} imposes the PPT condition, where $T_{A}$ denotes the partial transpose with respect to system $A$.

\section{Numerical Experiment Settings and Additional Results}

This section details the numerical experiment settings referenced in the main text. We also provide additional results and figures to substantiate our findings.

Riemannian optimizations were performed using the Riemannian conjugate gradient solver provided by \texttt{manopt} package in MATLAB, while SDPs for PPT relaxations were solved using the \texttt{SeDuMi} solver provided by \texttt{cvx} package. Source code is available upon request.

\subsection{Maximizing the Distillation Fidelity}

We apply the proposed framework to optimize both the average fidelity and the probabilistic fidelity (where a non-zero failure probability is allowed to enhance the final fidelity). We present achievable bounds for these quantities under various LOCC schemes. The optimization results include suboptimal objective values and the corresponding implementable protocols.

Consider $N$ agents sharing $M$ copies of a noisy maximally entangled state (MES) $\rho_{{[N]}}^{\{k\}} \in \mathrm{\cB}(\cH_{1}^{\{k\}}\otimes \cdots \otimes \cH_{N}^{\{k\}})$, for $k = 1,\dots,M$. The reduced Hilbert space for agent $X$ is denoted by $\cH_{X} = \cH_{X}^{\{1\}}\otimes \cdots \otimes \cH_{X}^{\{M\}}$. The goal of entanglement distillation is to prepare the maximally entangled state $\Phi_{[N]}^{\{1\}} = \ketbra{\Phi^{+}}{\Phi^{+}}_{[N]}^{\{1\}}$ in the space $\cB(\cH_{[N]}^{\{1\}})$ following a fixed-round LOCC protocol.

We consider three noise models: the depolarizing channel (depo.), the amplitude-damping channel (a.d.), and the dephasing channel (deph.), parameterized by $\gamma_d$, $\gamma_a$, and $\gamma_p$, respectively. Their actions on a state $\Phi$ are defined as:
\begin{align}
    &\cN_{\mathrm{depo.}}(\gamma_d, \Phi) = (1-\gamma_d)\Phi ~+~ \gamma_d \idop/d,\\
    &\cN_{\mathrm{a.d.}}(\gamma_a, \Phi) \!= \!K_{0} \Phi K_{0}^\dagger \!+\!\! \sum_{i=1}^{d-1}\!\gamma_a\ketbra{i\!-\!1}{i}\Phi\ketbra{i}{i\!-\!1},\\
    &\cN_{\mathrm{deph.}}(\gamma_p, \Phi) = \gamma_p \hat{\Phi} + (1-\gamma_p)\Phi,
\end{align}
where $d$ is the dimension of $\Phi$, $K_{0}=\ketbra{0}{0} + \sum_{i=1}^{d-1}\sqrt{1-\gamma_a}\ketbra{i}{i}$, and $\hat{\Phi}$ denotes a copy of $\Phi$ with all off-diagonal elements set to zero. In the case of 2-copy non-\textit{i.i.d.} inputs, the first and second copies are subjected to amplitude-damping and depolarizing noise, respectively.

Let $\cE_{\bm{j}}$ denote the CP map corresponding to the measurement outcome sequence $\bm{j}$. The average distillation fidelity is defined as
\begin{align}
    \cF_{\rm ave} = \sum_{\bm{j}} \Tr[\cE_{\bm{j}}(\rho^{\{[M]\}}_{[N]})\Phi_{[N]}^{\{1\}}],
\end{align}
where the summation runs over the set of outcomes indicating a successful protocol. For the optimization of average distillation fidelity, we set the instrument order to $S=2$ and the Kraus order to $T=1$.

\begin{figure}[h]
    \centering
    % 第一行两张图
    \includegraphics[width=0.35\textwidth]{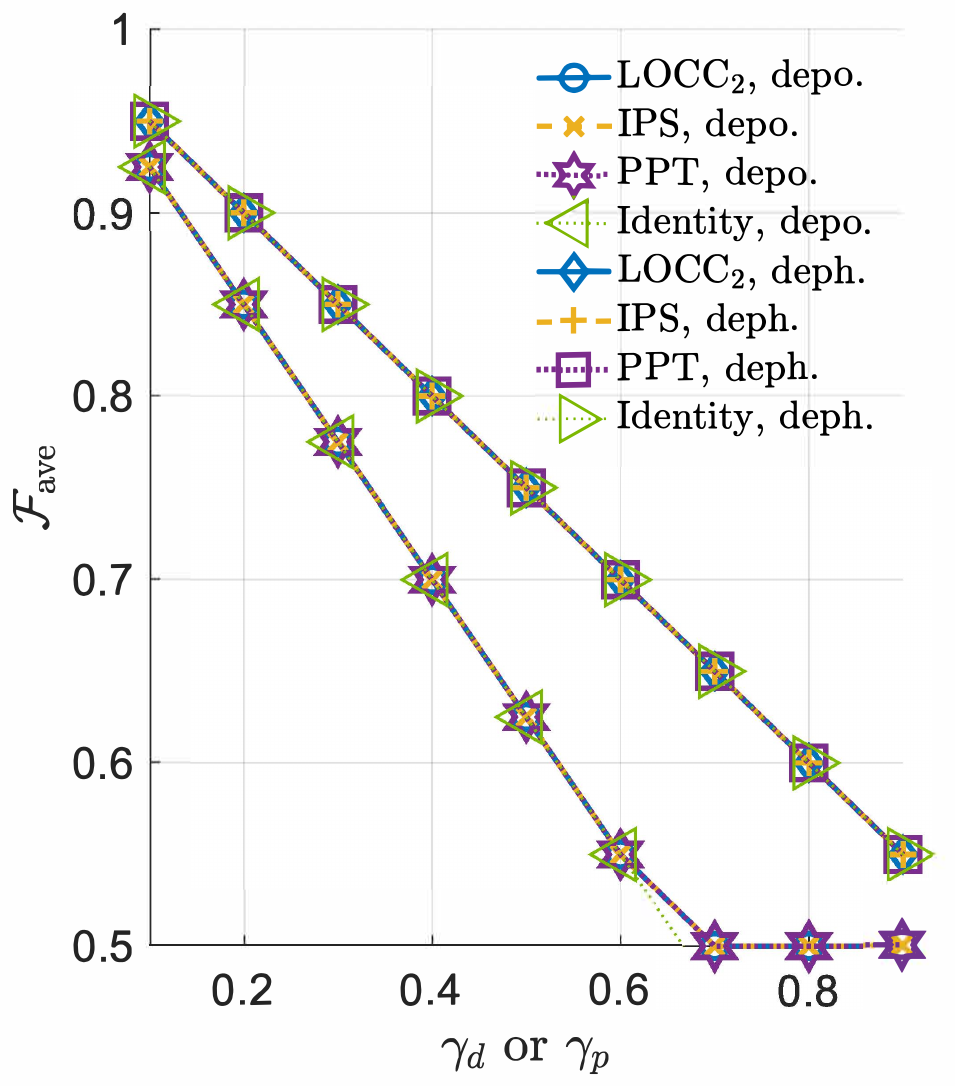}
    \caption{Optimized average distillation fidelity from MESs under depolarizing and dephasing noise via IPS, $\locc_1$, and $\locc_2$. In this setting, all PPT and LOCC schemes fail to distill entanglement effectively.}
    \label{fig:depo_deph_fid_res}
\end{figure}

Figure~\ref{fig:depo_deph_fid_res} presents the distillation results for noisy MESs under depolarizing and dephasing noise, complementing the findings in the main text. We observe no significant difference in average fidelity between $\locc_2$, IPS, PPT, and identity operations, indicating that entanglement cannot be successfully distilled in this specific regime.

\begin{figure}[h]
    \centering
    % 第一行两张图
    \includegraphics[width=.66\textwidth]{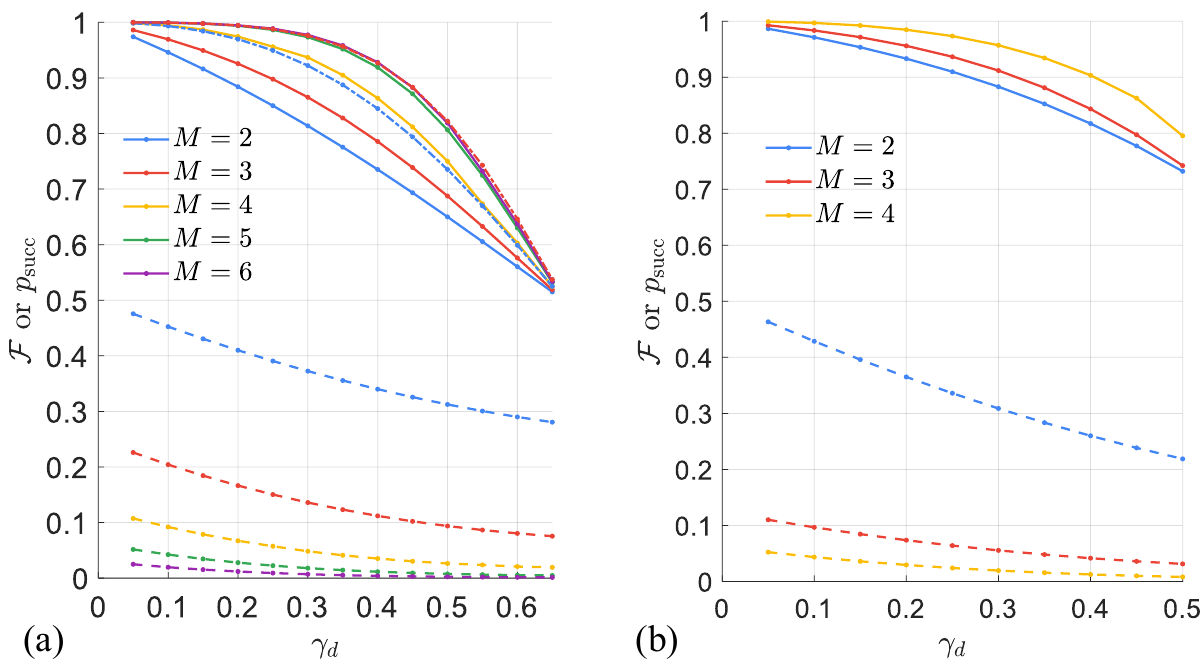}
    \caption{Optimization results for the distillation fidelity of a single outcome via CMPS. The solid and dashed lines represent the suboptimal fidelity and the success probability, respectively. Panels (a) and (b) display the suboptimal fidelity for bipartite and tripartite distillation, respectively, where the noise is modeled as a depolarizing channel with parameter $\gamma_d$ and $M$ denotes the number of copies. Compared to Fig.~\ref{fig:depo_deph_fid_res}, higher fidelity is achieved at the cost of a lower success probability. These results further demonstrate the scalability of our framework in optimizing multi-copy systems.}
    \label{fig:depo_res_multi_cps}
\end{figure}

To achieve a higher final fidelity by allowing for a non-zero probability of failure, we optimize the distillation fidelity conditioned on a specific instrument outcome sequence $\bm{0}$. The distillation fidelity is defined as
\begin{align}\label{eq:dstl_fid}
    \cF = \frac{\Tr[\cE_{\bm{0}}(\rho^{\{[M]\}}_{[N]}) \Phi_{[N]}^{\{1\}}]}{\Tr[\cE_{\bm{0}}(\rho^{\{[M]\}}_{[N]})]},
\end{align}
where ${\cE_{\bm{0}}(\rho^{\{[M]\}}_{[N]})}/{\Tr[\cE_{\bm{0}}(\rho^{\{[M]\}}_{[N]})]}$ represents the normalized output state. 

We illustrate the numerical lower bounds for the maximal fidelity of a single outcome via the CMPS scheme in Fig.~2(c) of the main text and Fig.~\ref{fig:depo_res_multi_cps} of this Supplemental Material. The POVM operator, when the protocol declares success, is fixed as $\ketbra{0}{0}$ for all systems except the first copy. Figure~\ref{fig:depo_res_multi_cps} further illustrates the optimized fidelity and success probability for bipartite and tripartite distillation in multi-copy systems, substantiating the scalability claims made in the main text.

For the running time comparison shown in Fig.~3 of the main text, the instrument order and Kraus order were set to $S = 2$ and $T=4$, respectively.

\subsection{Improved Achievable Bound of Two-way Distillable Entanglement}

Recall that the regularized formula for two-way distillable entanglement is given by
\begin{align}
    D(\rho_{AB})=\lim_{n\to \infty}\frac{1}{n}D^{(1)}(\rho^{\otimes n}_{AB}),
\end{align}
where
\begin{align}
    D^{(1)}(\rho_{AB}) = \max_{\cV} I(A'\rangle B')_{\cV(\rho_{AB})}.
\end{align}
Here, $I(A\rangle B)_{\rho_{AB}}=S(\rho_B)-S(\rho_{AB})$ denotes the coherent information, $S(\rho)$ is the von Neumann entropy, and $\cV: AB\to A'B'$ represents an $\locc_2$ operation \cite{devetak_distillation_2005}. An achievable bound for two-way distillable entanglement can thus be obtained by optimizing the block-length coherent information for a finite $n$.

We employ a simplified $\locc_2$ scheme to derive this bound, where all intermediate CPTP channels are restricted to the identity map. The Kraus order is set to $T=1$. For the instrument order, we use $S=2$ for \textit{i.i.d.} inputs and $S=4$ for non-\textit{i.i.d.} inputs. The input state is the Choi state of the generalized amplitude damping channel (GADC), $\rho_{AB}=\cI\otimes\cN_{g.a.d.}(\Phi_{AB})$, with Kraus operators given by:
\begin{align}
    K_1 =& \sqrt{1-\gamma_n}(\ketbra{0}{0}+\sqrt{1-\gamma_a}\ketbra{1}{1}),\\
    K_2 =& \sqrt{\gamma_a(1-\gamma_n)}(\ketbra{0}{1}),\\
    K_3 =& \sqrt{\gamma_n}(\sqrt{1-\gamma_a}\ketbra{0}{0}+\ketbra{1}{1}),\\
    K_4 =& \sqrt{\gamma_a\gamma_n}(\ketbra{1}{0}),
\end{align}
parameterized by $\gamma_a,\gamma_n\in[0,1]$. Note that when $\gamma_n=0$, the GADC reduces to the standard amplitude damping channel.

\subsection{State Merging}
\label{appendix:state_merging}

State merging is a canonical LOCC primitive for transferring Alice's share of a bipartite state to Bob while preserving correlations with an inaccessible reference system \cite{horodecki_partial_2005,horodecki_quantum_2006}. We use this task as a complementary test of the product-Stiefel framework because the objective differs from entanglement distillation: the protocol must reproduce a target joint state on Bob's side, and the finite-round one-shot fidelity is not specified by the conditional entropy $S(A|B)$ alone. The framework therefore applies by retaining the same Stiefel-parameterized LOCC protocol class and replacing the distillation overlap by the state-merging fidelity.

Let $\psi_{RAB}$ denote the tripartite pure input state, where $R$ is the reference system. Alice and Bob share a prior MES $\Phi_{A_eB_e}$ with Schmidt rank $k$, and the protocol may output a target MES $\Phi'_{A_e'B_e'}$ with Schmidt rank $m$. The objective is to produce an output state $\rho_{R A_e'B_e'B'B''}$ with maximal fidelity to $\Omega_{\rm mer}=\Phi'_{A_e'B_e'} \otimes \psi_{RB'B''}$, where $\psi_{RB'B''} = \Tr_{AB}[\Pi_{AB\leftrightarrow B'B''}(\psi_{RAB}\otimes \idop_{B'B''})]$ and $\Pi_{AB\leftrightarrow B'B''}$ denotes the SWAP operation between $AB$ and $B'B''$. Alice and Bob are restricted to local operations on $AA_eA_e'$ and $BB_eB_e'B'B''$, respectively, assisted by classical communication.

For an LOCC protocol $\mathcal{N}_{\mathbf{K}}$ parameterized by $\mathbf{K} \in \mathcal{M}$ and a success set $\mathcal{S}$ of outcome sequences, the optimized merging fidelity is
\begin{align}
    \cF_{\rm mer}^*(\mathcal{S}) =
    \max_{\mathbf{K} \in \mathcal{M}}
    \frac{\sum_{\bm{j}\in\mathcal{S}}\Tr[\rho_{\bm{j}}(\mathbf{K})\,\Omega_{\rm mer}]}
    {\sum_{\bm{j}\in\mathcal{S}}\Tr[\rho_{\bm{j}}(\mathbf{K})]},
\end{align}
where $\rho_{\bm{j}}(\mathbf{K})=\mathcal{N}_{\mathbf{K}}(\psi_{RAB}\Phi_{A_eB_e})\vert_{\bm{j}}$ is the sub-normalized output conditioned on outcome $\bm{j}$. If $\mathcal{S}$ contains all outcomes, this objective reduces to the average merging fidelity; if $\mathcal{S}$ is a designated success set, the numerator is normalized by the total success probability. As in the main text, this is an unconstrained Riemannian optimization over the product Stiefel manifold $\mathcal{M} \in \{\mathcal{M}_{\locc_r}, \mathcal{M}_{\mathrm{IPS}}, \mathcal{M}_{\mathrm{CMPS}}\}$.

\begin{figure}
    \centering
    \includegraphics[width=0.3\linewidth]{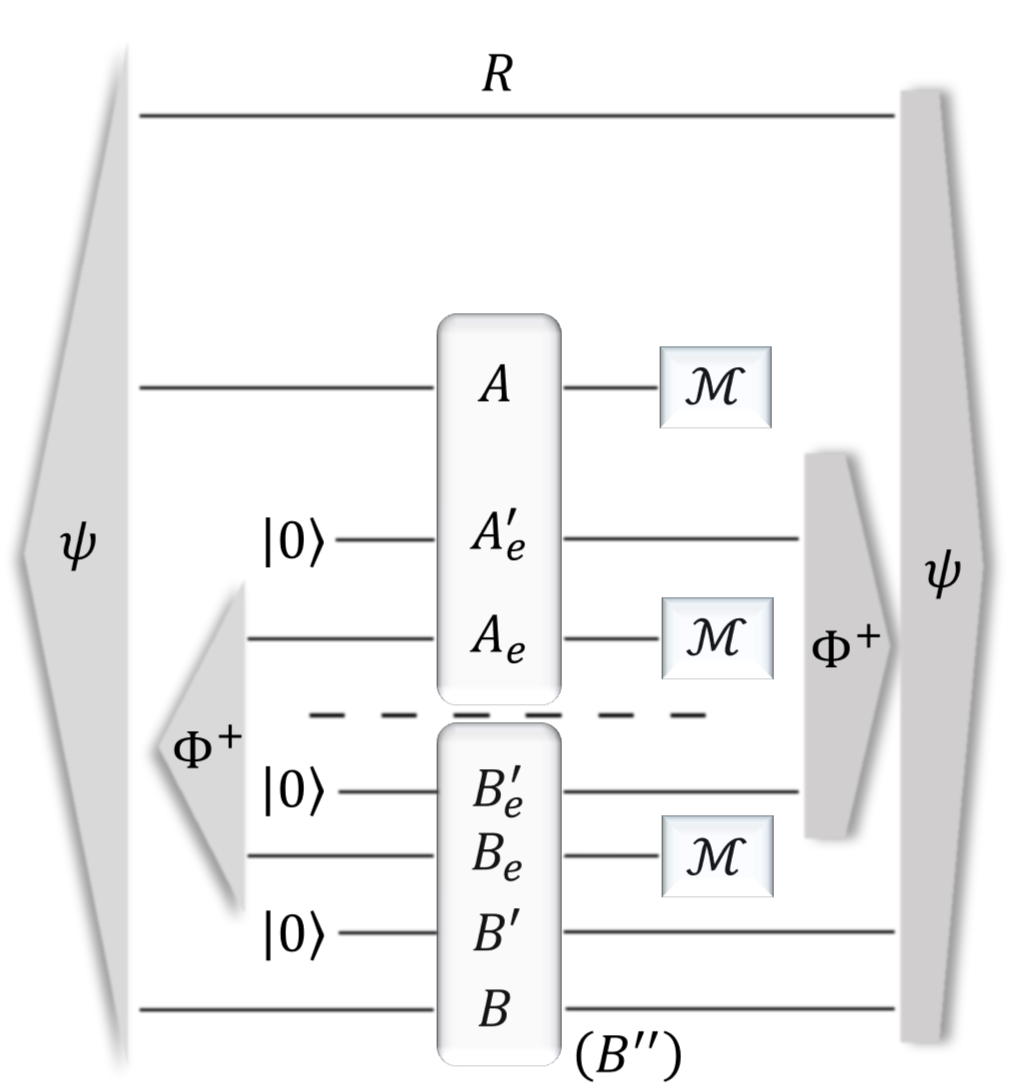}
    \caption{Schematic diagram of CMPS-based state merging. Rounded rectangles represent CPTP channels, while rectangles marked by $\cM$ represent the specified POVM operators. Output systems $A_e'$, $B_e'$, and $B'$ are initialized to $\ket{0}$. The output system $B''$ is identified with $B$.}
    \label{fig:CMPS_based_merging}
\end{figure}

In the numerical implementation, the dimensions of systems $R$, $A$, $B$, $B'$, and $B''$ are set to 2. The LOCC protocols are implemented via the CMPS scheme, as illustrated in Fig.~\ref{fig:CMPS_based_merging}. The output systems $A_e'$, $B_e'$, and $B'$ are initialized to $\ket{0}$, while $B''$ is identified with $B$. Alice and Bob independently apply local channels to $AA_eA_e'$ and $B B_e B_e' B'$, respectively. A specified POVM $\cM$ is then performed on systems $A$, $A_e$, and $B_e$, yielding outcomes $\bm{j} = j_Aj_{A_e}j_{B_e}$. In the success-branch setting, the target outcome sequence is $\bm{0}$, and the optimized fidelity is
\begin{align}
    \cF_{\rm mer} = \frac{\Tr[\cE_{\bm{0}}(\psi_{RAB}\Phi_{A_eB_e}) \Omega_{\rm mer}]}{\Tr[\cE_{\bm{0}}(\psi_{RAB}\Phi_{A_eB_e})]}.
\end{align}
For the deterministic setting, the corresponding average merging fidelity is
\begin{align}
    \cF_{\rm mer,ave} = \sum_{\bm{j}} {\Tr[\cE_{\bm{j}}(\psi_{RAB}\Phi_{A_eB_e}) \Omega_{\rm mer}]},
\end{align}
where the summation extends over all outcome sequences $\bm{j}$.

We investigate three resource settings. The baseline case $k=1$ and $m=1$ probes one-shot merging without entanglement assistance or output entanglement. The assisted case $k=2$ and $m=1$ tests whether one consumed e-bit suffices for perfect qubit-state merging. The catalytic case $k=2$ and $m=2$ tests whether a supplied Bell state can enlarge the achievable region when it must be recovered at the output.

We randomly sampled 20,000 pure states $\psi_{RAB}$ according to the Haar measure. For each sampled state, we computed the conditional entropy $S(A|B)=S(\psi_{AB})-S(\psi_B)$ and optimized the corresponding merging fidelity. All CMPS optimizations for state merging employ the full Kraus order. For maximizing $\cF_{\rm mer}$, success is declared when the POVM outcome corresponds to $\ketbra{0}{0}$. For optimizing $\cF_{\rm mer,ave}$, the POVM operator corresponds to the identity $\idop$.

\begin{figure*}
    \centering
    \includegraphics[width=.95\linewidth]{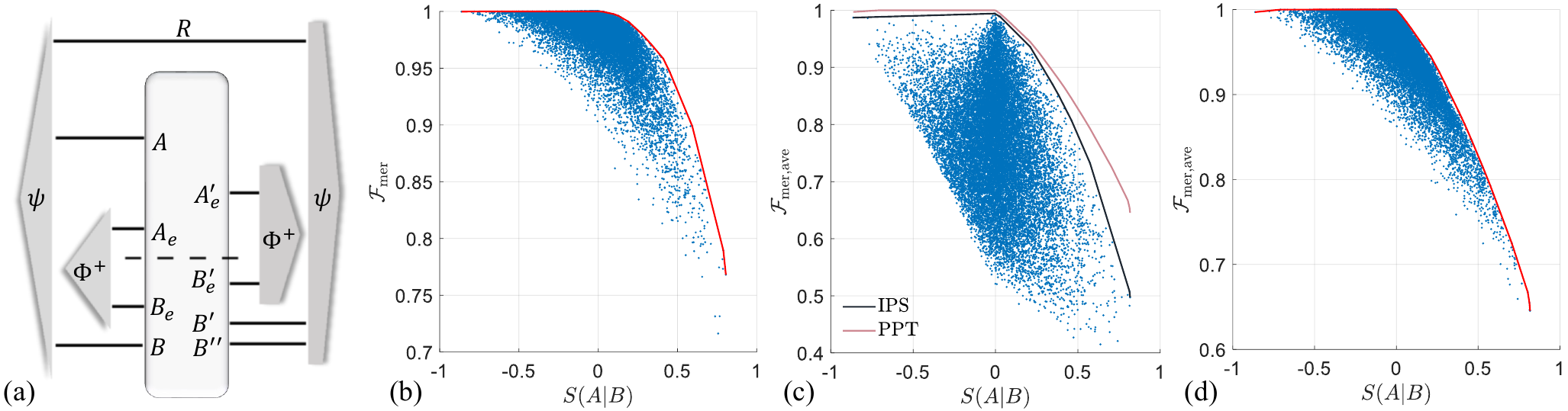}
    \caption{One-shot state merging analysis.
(a) Schematic of the protocol. The central rounded rectangle represents the quantum operation CMPS or PPT. The dashed line separates the parties of Alice ($A$) and Bob ($B$).
(b)--(d) Results for 20,000 Haar-random states in the baseline scenario ($k=1, m=1$). Scatter dots map out the achievable regions.
(b) shows the achievable region and upper envelope of the success-branch fidelity $\cF_{\rm mer}$ via CMPS, while (c) and (d) plot the corresponding average-fidelity regions $\cF_{\rm mer, ave}$ via CMPS and PPT operations, respectively.
To visualize the operational gap, the upper envelope of the PPT region from (d) is superimposed in (c).
The envelopes provide numerical estimates of the maximal fidelity function $f(x) = \{\max_{\psi_{RAB}}\cF : S(A|B)_{\psi_{RAB}}=x\}$. The differences in geometric shape and upper envelope between CMPS and PPT suggest a numerical operational gap between the two schemes.}
    \label{fig:merging_res}
\end{figure*}

\subsubsection{Results}

We first analyze the baseline scenario ($k=1, m=1$), corresponding to merging without entanglement consumption or generation. Fig.~\ref{fig:merging_res}(b) illustrates the achievable region found for the CMPS scheme in this baseline setting and delineates its numerical upper envelope. Evaluating each sampled state across extensive random initializations mitigates local optima, yielding a robustly mapped achievable region for the tested instances. The profile of this envelope suggests that the best observed fidelity follows a monotonic trend with respect to the conditional entropy $S(A|B)$. Consequently, this numerical envelope provides an estimated performance frontier of the CMPS scheme.

Fig.~\ref{fig:merging_res}(c,d) map out the achievable regions for the average merging fidelity. We observe that the geometric shape of the numerically achievable region obtained via CMPS differs from that obtained by PPT operations. In particular, the upper-envelope difference indicates a numerical operational gap between the two schemes. While PPT bounds serve as theoretical upper limits, our results suggest that they can overestimate the fidelity accessible to an explicit LOCC sub-scheme, highlighting the practical value of direct protocol optimization for realistic protocol design.

To probe the resource requirements for perfect merging, we introduce a single e-bit of pre-shared entanglement ($k=2$, $m=1$). This setting represents the consumption of one e-bit to assist transmission. Our optimization finds achievable fidelities approaching $1$ for all sampled states across the entire entropy spectrum, suggesting that one e-bit may be sufficient for complete one-shot merging in the sampled qubit systems considered here. We also investigate the catalytic setting ($k=2$, $m=2$), where a Bell pair is supplied to assist the protocol but must be recovered at the output. The achievable region for this catalytic case coincides numerically with that of the unassisted regime ($k=1$, $m=1$), indicating that no expansion from simple Bell-state catalysis was observed for the tested one-shot CMPS merging protocols.

\section{Gradient-Based Optimization on the Product Stiefel Manifold}
\label{appendix:stiefel_manifold}

The Stiefel manifold $\mathrm{St}(n,p)$ is defined as the set of $p$ orthonormal vectors in $\mathbb{C}^{n}$ (or $\mathbb{R}^{n}$ for real cases), representing an embedded submanifold of $\mathbb{C}^{n\times p}$~\cite{boumal_introduction_2023}. An element $X \in \mathrm{St}(n,p)$ satisfies the orthogonality constraint $X^{\dagger} X = I_p$.

Optimization on the Stiefel manifold generalizes standard unconstrained optimization by adhering to the manifold geometry. The core procedure involves three steps: (i) defining a metric and the corresponding tangent space $\mathcal{T}_X \mathrm{St}(n,p)$; (ii) projecting the Euclidean gradient onto the tangent space to obtain the Riemannian gradient; and (iii) updating the variable along a geodesic-like curve via a retraction mapping.

The tangent space $\mathcal{T}_X \mathcal{M}$ at a point $X$ is formally defined as the vector space of velocities for all smooth curves passing through $X$:
\begin{align}
    \mathcal{T}_X \mathcal{M} = \left\{ \dot{R}(0) \left| \begin{gathered}
        R: \mathcal{I} \to \mathcal{M} \text{ is smooth}, \\
        R(0) = X
    \end{gathered}\right.\right\}.
\end{align}

Intuitively, this space is the orthogonal complement to the gradients of the defining constraints. Since the manifold is a level set of the constraints, any motion restricted to the manifold must be perpendicular to the normal directions of the constraints.
For the Stiefel manifold defined by $X^\dagger X = I_p$, differentiating this condition along a curve yields $\dot{R}^\dagger R + R^\dagger \dot{R} = 0$. Consequently, the tangent space is the set of matrices satisfying the skew-Hermitian condition:
\begin{align}
    \mathcal{T}_X \mathrm{St}(n,p) = \{Z \in \mathbb{C}^{n \times p} \mid Z^\dagger X + X^\dagger Z = 0\}.
\end{align}

To perform gradient descent, we define the Riemannian metric using the canonical Euclidean inner product, $\langle Z_1, Z_2 \rangle = \operatorname{Re}\operatorname{Tr}(Z_1^\dagger Z_2)$. Given a scalar objective function $f(X)$, the {Euclidean gradient} is given by $G = \nabla f(X) = \frac{\partial f}{\partial X^*}$. The {Riemannian gradient} $U \in \mathcal{T}_X \mathrm{St}(n,p)$ is obtained by projecting $G$ onto the tangent space:
\begin{align}
  U = \operatorname{grad} f(X) = G - \frac{1}{2} X (X^\dagger G + G^\dagger X).
\end{align}

Moving along the tangent vector $U$ would generally violate the manifold constraints. To ensure the updated point remains on $\mathrm{St}(n,p)$, we employ a {retraction} mapping $R_X: \mathcal{T}_X \mathrm{St}(n,p) \to \mathrm{St}(n,p)$. A computationally efficient choice is the QR-decomposition-based retraction:
\begin{align}
  R_X(\xi) = \operatorname{qf}(X + \xi),
\end{align}
where $\operatorname{qf}(A)$ denotes the $Q$ factor of the QR decomposition of $A$, adjusted to ensure specific diagonal elements of $R$ are positive for uniqueness.

The Riemannian gradient descent algorithm for a product of Stiefel manifolds $\mathcal{M} = \mathrm{St}(n_1, p_1) \times \dots \times \mathrm{St}(n_k, p_k)$ applies these operations component-wise. A single iteration with step size $\eta$ proceeds as follows:
\begin{enumerate}
    \item \textbf{Euclidean Gradient:} Compute $G = \nabla f(X)$ in the ambient space.
    \item \textbf{Riemannian Gradient:} Compute the tangent projection $U = G - \frac{1}{2} X (X^\dagger G + G^\dagger X)$.
    \item \textbf{Update:} Update the parameter via retraction: $X_{k+1} = \operatorname{qf}(X_k - \eta U)$.
\end{enumerate}

\bibliography{ref}

%% file: pretex.tex
\usepackage[dvipsnames]{xcolor}
\usepackage{framed}
\definecolor{shadecolor}{rgb}{0.9,0.9,0.9}

\usepackage{mathtools}
\usepackage{amsmath}
\usepackage[shortlabels]{enumitem}
\usepackage{graphicx,epic,eepic,epsfig,amsmath,latexsym,amssymb,verbatim,color}
\usepackage{bm}
\usepackage{subfigure}
\usepackage{amsfonts}       % blackboard math symbols
\usepackage{nicefrac}       % compact symbols for 1/2, etc.
\usepackage{dsfont}       % blackboard math symbols

\usepackage{amsmath}
\usepackage{bbm}

\usepackage{float}
\usepackage{tikz}
\usetikzlibrary{chains}
\usetikzlibrary{fit}
\usepackage{pgflibraryarrows}		%optional
\usepackage{pgflibrarysnakes}		%optional

\usepackage{epsfig}
\usetikzlibrary{shapes.symbols,patterns} % for source symbols
\usepackage{pgfplots}

\usepackage[strict]{changepage}
\usepackage{hyperref}
\hypersetup{colorlinks=true,citecolor=blue,linkcolor=blue,filecolor=blue,urlcolor=blue,breaklinks=true}

\usepackage[marginal]{footmisc}
\usepackage{url}
\usepackage{theorem}

\newtheorem{proposition}{Proposition}

\def\squareforqed{\hbox{\rlap{$\sqcap$}$\sqcup$}}
\def\qed{\ifmmode\squareforqed\else{\unskip\nobreak\hfil
\penalty50\hskip1em\null\nobreak\hfil\squareforqed
\parfillskip=0pt\finalhyphendemerits=0\endgraf}\fi}
\def\endenv{\ifmmode\;\else{\unskip\nobreak\hfil
\penalty50\hskip1em\null\nobreak\hfil\;
\parfillskip=0pt\finalhyphendemerits=0\endgraf}\fi}

\newcounter{remark}
\newenvironment{remark}[1][]{\refstepcounter{remark}\par\medskip\noindent%
\textbf{Remark~\theremark #1} }{\medskip}

\newcounter{example}

\mathchardef\ordinarycolon\mathcode`\:
\mathcode`\:=\string"8000
\def\vcentcolon{\mathrel{\mathop\ordinarycolon}}
\begingroup \catcode`\:=\active
  \lowercase{\endgroup
  \let :\vcentcolon
  }

\usepackage{cleveref}
\usepackage{graphicx}
\usepackage{xcolor}

\RequirePackage[framemethod=default]{mdframed}
\newmdenv[skipabove=7pt,
skipbelow=7pt,
backgroundcolor=darkblue!15,
innerleftmargin=5pt,
innerrightmargin=5pt,
innertopmargin=5pt,
leftmargin=0cm,
rightmargin=0cm,
innerbottommargin=5pt,
linewidth=1pt]{tBox}

\newmdenv[skipabove=7pt,
skipbelow=7pt,
backgroundcolor=red!15,
innerleftmargin=5pt,
innerrightmargin=5pt,
innertopmargin=5pt,
leftmargin=0cm,
rightmargin=0cm,
innerbottommargin=5pt,
linewidth=1pt]{rBox}

\newmdenv[skipabove=7pt,
skipbelow=7pt,
backgroundcolor=blue2!25,
innerleftmargin=5pt,
innerrightmargin=5pt,
innertopmargin=5pt,
leftmargin=0cm,
rightmargin=0cm,
innerbottommargin=5pt,
linewidth=1pt]{dBox}
\newmdenv[skipabove=7pt,
skipbelow=7pt,
backgroundcolor=darkkblue!15,
innerleftmargin=5pt,
innerrightmargin=5pt,
innertopmargin=5pt,
leftmargin=0cm,
rightmargin=0cm,
innerbottommargin=5pt,
linewidth=1pt]{sBox}
\definecolor{darkblue}{RGB}{0,76,156}
\definecolor{darkkblue}{RGB}{0,0,153}
\definecolor{blue2}{RGB}{102,178,255}
\definecolor{darkred}{RGB}{195,0,0}
\newcommand{\nc}{\newcommand}
\nc{\rnc}{\renewcommand}
\nc{\lbar}[1]{\overline{#1}}
\nc{\bra}[1]{\langle#1|}
\nc{\ket}[1]{|#1\rangle}
\nc{\ketbra}[2]{|#1\rangle\!\langle#2|}
\nc{\braket}[2]{\langle#1|#2\rangle}

\nc{\idop}{{\mathds 1}}
\nc{\proj}[1]{| #1\rangle\!\langle #1 |}
\nc{\avg}[1]{\langle#1\rangle}
\nc{\smfrac}[2]{\mbox{$\frac{#1}{#2}$}}
\nc{\tr}{\operatorname{Tr}}
\nc{\ox}{\otimes}
\nc{\dg}{\dagger}
\nc{\dn}{\downarrow}
\nc{\cA}{{\cal A}}
\nc{\cB}{{\cal B}}
\nc{\cC}{{\cal C}}
\nc{\cD}{{\cal D}}
\nc{\cE}{{\cal E}}
\nc{\cF}{{\cal F}}
\nc{\cG}{{\cal G}}
\nc{\cH}{{\cal H}}
\nc{\cI}{{\cal I}}
\nc{\cJ}{{\cal J}}
\nc{\cK}{{\cal K}}
\nc{\cL}{{\cal L}}
\nc{\cM}{{\cal M}}
\nc{\cN}{{\cal N}}
\nc{\cO}{{\cal O}}
\nc{\cP}{{\cal P}}
\nc{\cQ}{{\cal Q}}
\nc{\cR}{{\cal R}}
\nc{\cS}{{\cal S}}
\nc{\cT}{{\cal T}}
\nc{\cU}{{\cal U}}
\nc{\cV}{{\cal V}}
\nc{\cX}{{\cal X}}
\nc{\cY}{{\cal Y}}
\nc{\cZ}{{\cal Z}}
\nc{\cW}{{\cal W}}
\nc{\st}{{\mathrm{St}}}

\nc{\csupp}{{\operatorname{csupp}}}
\nc{\qsupp}{{\operatorname{qsupp}}}
\nc{\var}{{\operatorname{var}}}
\nc{\rar}{\rightarrow}
\nc{\lrar}{\longrightarrow}
\nc{\polylog}{{\operatorname{polylog}}}
\nc{\wt}{{\operatorname{wt}}}
\nc{\av}[1]{{\left\langle {#1} \right\rangle}}
\nc{\supp}{{\operatorname{supp}}}

\nc{\argmin}{{\operatorname{argmin}}}

\def\x{\xi}

\nc{\RR}{{{\mathbb R}}}
\nc{\CC}{{{\mathbb C}}}
\nc{\FF}{{{\mathbb F}}}
\nc{\NN}{{{\mathbb N}}}
\nc{\ZZ}{{{\mathbb Z}}}
\nc{\PP}{{{\mathbb P}}}
\nc{\QQ}{{{\mathbb Q}}}
\nc{\JJ}{{{\mathbb J}}}
\nc{\UU}{{{\mathbb U}}}
\nc{\EE}{{{\mathbb E}}}
\nc{\TT}{{{\mathbb T}}}
\nc{\id}{{\operatorname{id}}}

\nc{\CHSH}{{\operatorname{CHSH}}}

\nc{\be}{\begin{equation}}
\nc{\ee}{{\end{equation}}}
\nc{\bea}{\begin{eqnarray}}
\nc{\eea}{\end{eqnarray}}
\nc{\<}{\langle}
\rnc{\>}{\rangle}
\nc{\rU}{\mbox{U}}

\nc{\ob}[1]{#1}

\nc{\SEP}{{\text{\rm SEP}}}
\nc{\NS}{{\text{\rm NS}}}
\nc{\LOCC}{{\text{\rm LOCC}}}
\nc{\PPT}{{\text{\rm PPT}}}
\nc{\EXT}{{\text{\rm EXT}}}
\nc{\Sym}{{\operatorname{Sym}}}

\nc{\ERLO}{{E_{\text{r,LO}}}}
\nc{\ERLOCC}{{E_{\text{r,LOCC}}}}
\nc{\ERPPT}{{E_{\text{r,PPT}}}}
\nc{\ERLOCCinfty}{{E^{\infty}_{\text{r,LOCC}}}}
\nc{\Aram}{{\operatorname{\sf A}}}

\usepackage{tikz}
\usepackage{hyperref}
\hypersetup{colorlinks=true,citecolor=blue,linkcolor=blue,filecolor=blue,urlcolor=blue,breaklinks=true}

\makeatletter
\def\grd@save@target#1{%
  \def\grd@target{#1}}
\def\grd@save@start#1{%
  \def\grd@start{#1}}
\tikzset{
  grid with coordinates/.style={
    to path={%
      \pgfextra{%
        \edef\grd@@target{(\tikztotarget)}%
        \tikz@scan@one@point\grd@save@target\grd@@target\relax
        \edef\grd@@start{(\tikztostart)}%
        \tikz@scan@one@point\grd@save@start\grd@@start\relax
        \draw[minor help lines,magenta] (\tikztostart) grid (\tikztotarget);
        \draw[major help lines] (\tikztostart) grid (\tikztotarget);
        \grd@start
        \pgfmathsetmacro{\grd@xa}{\the\pgf@x/1cm}
        \pgfmathsetmacro{\grd@ya}{\the\pgf@y/1cm}
        \grd@target
        \pgfmathsetmacro{\grd@xb}{\the\pgf@x/1cm}
        \pgfmathsetmacro{\grd@yb}{\the\pgf@y/1cm}
        \pgfmathsetmacro{\grd@xc}{\grd@xa + \pgfkeysvalueof{/tikz/grid with coordinates/major step}}
        \pgfmathsetmacro{\grd@yc}{\grd@ya + \pgfkeysvalueof{/tikz/grid with coordinates/major step}}
        \foreach \x in {\grd@xa,\grd@xc,...,\grd@xb}
        \node[anchor=north] at (\x,\grd@ya) {\pgfmathprintnumber{\x}};
        \foreach \y in {\grd@ya,\grd@yc,...,\grd@yb}
        \node[anchor=east] at (\grd@xa,\y) {\pgfmathprintnumber{\y}};
      }
    }
  },
  minor help lines/.style={
    help lines,
    step=\pgfkeysvalueof{/tikz/grid with coordinates/minor step}
  },
  major help lines/.style={
    help lines,
    line width=\pgfkeysvalueof{/tikz/grid with coordinates/major line width},
    step=\pgfkeysvalueof{/tikz/grid with coordinates/major step}
  },
  grid with coordinates/.cd,
  minor step/.initial=.2,
  major step/.initial=1,
  major line width/.initial=2pt,
}
\makeatother

\usepackage{thmtools}
\usepackage{thm-restate}
\usepackage{etoolbox}
\makeatletter
\def\problem@s{}
\newcounter{problems@cnt}

\newcommand{\allproblems}{\problem@s}
\makeatother

%% file: opdef.tex
\nc{\rank}{\operatorname{Rank}}

\nc{\Lin}{\operatorname{Lin}}
\nc{\Tr}{\operatorname{Tr}}

\nc{\tti}{\mathtt{i}}
\nc{\tto}{\mathtt{o}}
\nc{\tin}{\mathtt{in}}
\nc{\tout}{\mathtt{out}}

\nc{\rd}{{\rm d}}

\def\x{\xi}

\nc{\fJ}{{{\mathfrak J}}}
\nc{\VV}{{{\mathbb V}}}
\nc{\WW}{{{\mathbb W}}}
\nc{\KK}{{{\mathbb K}}}

\nc{\DD}{{{\mathbb D}}}